\documentclass[twoside,twocolumn,9pt]{article}
\usepackage{extsizes}
\usepackage[super,sort&compress,comma]{natbib} 
\usepackage[version=3]{mhchem}
\usepackage[left=1.5cm, right=1.5cm, top=1.785cm, bottom=2.0cm]{geometry}
\usepackage{balance}
\usepackage{enumitem}
\usepackage{mathptmx}
\usepackage{sectsty}
\usepackage{graphicx} 
\usepackage{lastpage}
\usepackage{comment}
\usepackage[format=plain,justification=justified,singlelinecheck=false,font={stretch=1.125,small,sf},labelfont=bf,labelsep=space]{caption}
\usepackage{float}
\usepackage{fancyhdr}
\usepackage{fnpos}
\usepackage[english]{babel}
\addto{\captionsenglish}{%
  
}
\usepackage{array}
\usepackage{droidsans}
\usepackage{charter}
\usepackage[T1]{fontenc}
\usepackage[usenames,dvipsnames]{xcolor}
\usepackage{setspace}
\usepackage[compact]{titlesec}
\usepackage{hyperref}
\usepackage{comment}
\usepackage[framemethod=TikZ]{mdframed}
\usepackage{xcolor}
\usepackage{xr}
\usepackage{comment}
\usepackage{minted}
\usepackage{siunitx}
\usepackage{amssymb}
\usepackage{tikz}
\usetikzlibrary{arrows.meta, positioning, shapes.geometric}
\DeclareSIUnit\torr{Torr}
\definecolor{darkred}{rgb}{0.6, 0.0, 0.0}
\definecolor{darkblue}{rgb}{0.0, 0.0, 0.6}
\definecolor{bg}{rgb}{0.95,0.95,0.95} 

\usepackage[most]{tcolorbox}
\tcbset{
  colback=blue!3!white,
  colframe=blue!50!black,
  boxrule=0.6pt,
  arc=2pt,
  left=6pt,
  right=6pt,
  top=6pt,
  bottom=6pt
}

\definecolor{cream}{RGB}{222,217,201}

\begin{document}

\pagestyle{fancy}
\thispagestyle{plain}
\fancypagestyle{plain}{
\renewcommand{\headrulewidth}{0pt}
}

\makeFNbottom
\makeatletter
\renewcommand\LARGE{\@setfontsize\LARGE{15pt}{17}}
\renewcommand\Large{\@setfontsize\Large{12pt}{14}}
\renewcommand\large{\@setfontsize\large{10pt}{12}}
\renewcommand\footnotesize{\@setfontsize\footnotesize{7pt}{10}}
\makeatother

\renewcommand{\thefootnote}{\fnsymbol{footnote}}
\renewcommand\footnoterule{\vspace*{1pt}%
\color{cream}\hrule width 3.5in height 0.4pt \color{black}\vspace*{5pt}} 
\setcounter{secnumdepth}{5}

\makeatletter 
\renewcommand\@biblabel[1]{#1}            
\renewcommand\@makefntext[1]%
{\noindent\makebox[0pt][r]{\@thefnmark\,}#1}
\makeatother 
\renewcommand{\figurename}{\small{Fig.}~}
\sectionfont{\sffamily\Large}
\subsectionfont{\normalsize}
\subsubsectionfont{\bf}
\setstretch{1.125} 
\setlength{\skip\footins}{0.8cm}
\setlength{\footnotesep}{0.25cm}
\setlength{\jot}{10pt}
\titlespacing*{\section}{0pt}{4pt}{4pt}
\titlespacing*{\subsection}{0pt}{15pt}{1pt}

\fancyfoot{}
\fancyfoot[LO,RE]{\vspace{-7.1pt}\includegraphics[height=9pt]{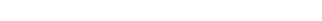}}
\fancyfoot[CO]{\vspace{-7.1pt}\hspace{13.2cm}\includegraphics{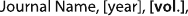}}
\fancyfoot[CE]{\vspace{-7.2pt}\hspace{-14.2cm}\includegraphics{head_foot/RF}}
\fancyfoot[RO]{\footnotesize{\sffamily{1--\pageref{LastPage} ~\textbar  \hspace{2pt}\thepage}}}
\fancyfoot[LE]{\footnotesize{\sffamily{\thepage~\textbar\hspace{3.45cm} 1--\pageref{LastPage}}}}
\fancyhead{}
\renewcommand{\headrulewidth}{0pt} 
\renewcommand{\footrulewidth}{0pt}
\setlength{\arrayrulewidth}{1pt}
\setlength{\columnsep}{6.5mm}
\setlength\bibsep{1pt}
\DeclareSIUnit\ppm{ppm}
\sisetup{
    range-phrase = {--},
    range-units = single
}

\makeatletter 
\newlength{\figrulesep} 
\setlength{\figrulesep}{0.5\textfloatsep} 

\newcommand{\topfigrule}{\vspace*{-1pt}%
\noindent{\color{cream}\rule[-\figrulesep]{\columnwidth}{1.5pt}} }

\newcommand{\botfigrule}{\vspace*{-2pt}%
\noindent{\color{cream}\rule[\figrulesep]{\columnwidth}{1.5pt}} }

\newcommand{\dblfigrule}{\vspace*{-1pt}%
\noindent{\color{cream}\rule[-\figrulesep]{\textwidth}{1.5pt}} }

\makeatother

\twocolumn[
  \begin{@twocolumnfalse}
\sffamily
\begin{tabular}{m{4.5cm} p{13.5cm} }

\includegraphics{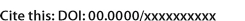} & \noindent\LARGE{\textbf{Autonomous Reliability Qualification of Ga\textsubscript{2}O\textsubscript{3}-based diode sensors via Safe Active Learning}} \\
\vspace{0.3cm} & \vspace{0.3cm} \\

 & \noindent\large{Davi Fébba, William A. Callahan, Anna Sacchi, and Andriy Zakutayev} \\

\includegraphics{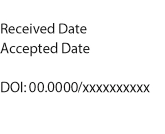} & \noindent\normalsize{Ultra-wide bandgap (UWBG) Ga\textsubscript{2}O\textsubscript{3} is a promising semiconductor for high-power and high-temperature electronics. Reliable qualification of these devices under extreme operating conditions is essential, yet conventional reliability testing is inherently time-consuming. Autonomous experimentation offers a new paradigm by enabling measurement planning and model refinement to evolve in parallel in real time. We present a Safe Active Learning (SAL) framework for autonomous reliability characterization of Ga\textsubscript{2}O\textsubscript{3}-based diode sensors under coupled thermal and hydrogen stress. We first evaluate SAL in simulation, where it safely expands the explored region while learning the evolving rectification surface. Second, we demonstrate SAL experimentally on an automated high-temperature probe-station platform using a Pt/Cr\textsubscript{2}O\textsubscript{3}:Mg/$\beta$-Ga\textsubscript{2}O\textsubscript{3} diode sensor of H\textsubscript{2} and temperature, spanning \SIrange{0}{800}{ppm} H\textsubscript{2} and \SIrange{350}{550}{\celsius}. Finally, we use the SAL-generated dataset for offline long-horizon forecasting of the diode current at a target voltage with a structured Gaussian-process model. Its condition-dependent Kohlrausch--Williams--Watts mean and residual covariance kernel were engineered with artificial-intelligence assistance using the SAL data and an auxiliary validation dataset spanning 1,000 hours at \SI{400}{\celsius} across multiple H\textsubscript{2} concentrations. This dataset guided kernel design and validation, and the resulting model captures its long-time, saturating degradation trends. Although demonstrated here for a rectifying Ga\textsubscript{2}O\textsubscript{3}-based diode, SAL is applicable to other device classes whenever a suitable safety observable can be measured in situ.}

\end{tabular}

 \end{@twocolumnfalse} \vspace{0.6cm}

  ]

\renewcommand*\rmdefault{bch}\normalfont\upshape
\rmfamily
\section*{}
\vspace{-1cm}


\footnotetext{Materials Science Center, National Laboratory of the Rockies (NLR), Golden, CO, 80401, USA. E-mail: DaviMarcelo.Febba@nlr.gov}
\footnotetext{\dag~Electronic Supplementary Information (ESI) available}



\section{Introduction}

The combination of a wide bandgap and exceptional breakdown field makes \(\beta\)-Ga\(_2\)O\(_3\) a strong candidate material for next-generation power electronics and applications in demanding conditions.\cite{pearton2018,he2024} However, the practical value of this material system is determined not only by its initial device performance, but also by its ability to preserve stable electrical behavior under sustained thermal, electrical, and chemical stress.\cite{wong_landscape_2023,lyle2022} For rectifying Ga\(_2\)O\(_3\)-based devices, reliability is closely tied to quantities such as rectification ratio, reverse leakage current, barrier integrity, and contact stability, all of which can evolve during operation and accelerated testing.\cite{wong_landscape_2023,he2024} This issue is especially important at elevated temperature and in hydrogen-containing ambients, where interfacial chemistry, defect-assisted conduction, and contact modification can significantly alter the measured current--voltage (IV) response.\cite{trinchi2004,heinselman2021,lee2010}

Traditional reliability characterization of rectifying semiconductor devices relies on repeated electrical measurements, including IV and capacitance--voltage (\(C\)\(V\)) characterization\cite{callahan2024a,heinselman2021,sun2023}, as a function of environmental conditions. Forward-bias stress studies have further shown that degradation can begin before obvious catastrophic failure, underscoring the need for continuous monitoring during long-duration experiments.\cite{sun2023} Although these methods are physically informative, they become increasingly inefficient when the relevant operating space is multidimensional and time dependent. 

Data-driven reliability modeling has emerged as an attractive complement to exhaustive empirical testing when degradation depends on many interacting variables and experiments are slow or costly. Across batteries and related energy systems, machine-learning approaches have shown that long-term degradation can often be inferred from early-cycle or partial-history data, enabling substantial reductions in test time while still preserving useful predictive accuracy.\cite{severson2019a,attia2020a,jiang2021a} Gaussian-process-based approaches have been especially attractive in degradation forecasting because they can model complex, nonlinear trajectories while quantifying uncertainty, including settings in which the model must generalize across operating conditions rather than interpolate only within a single fixed protocol.\cite{richardson2017a,richardson2019a} Neural-network-based methods have likewise become increasingly important in degradation prediction and remaining-useful-life estimation because they can capture highly nonlinear behavior in larger and more heterogeneous datasets.\cite{severson2019a,hachem2024a} 

However, conventional data-driven approaches rely on datasets collected under predetermined experimental protocols, so their fidelity is limited by the coverage of those datasets and they cannot adaptively direct measurements toward uncertain, highly informative, or safety-critical regions of the time-dependent operating space. The next step beyond data-driven prediction is autonomous experimentation, in which model updates and experimental decisions are coupled in a closed loop. 

In self-driving laboratories, automation, real-time analysis, and algorithmic experiment selection are integrated so that the platform can decide which condition to probe next with minimal human intervention. Recent advancements include autonomous electrochemical characterization of combinatorial thin-film proton-conducting oxide electrodes\cite{huang_benchtop_2026}, autonomous electrical measurements of thin-film material libraries\cite{thelen2023}, alongside a rapidly growing range of related applications across materials research.\cite{vogler_autonomous_2024,tom2024b} These advances suggest a compelling opportunity for semiconductor reliability characterization: rather than executing a predetermined matrix of stress conditions, an autonomous platform can adaptively select operating conditions based on the evolving state of the device and model uncertainty.

In this work, we apply this approach to autonomous reliability characterization of Ga\textsubscript{2}O\textsubscript{3}-based diode sensors under coupled thermal and hydrogen stress. Our goal is not simply to optimize device performance, but to learn how device response and rectifying behavior degrade over time while minimizing the risk of driving the device into destructive operating conditions. The main contributions of this work are threefold. First, we introduce a Safe Active Learning (SAL) algorithm for autonomous reliability characterization under time uncertainty. SAL uses the diode rectification ratio as a safety observable and models its evolution over time, temperature, and H\textsubscript{2} concentration with a Gaussian-process surrogate. It selects experiments using an adaptive completion-time window, a trust region anchored to previously verified safe conditions, and a two-phase strategy that transitions from conservative exploration to controlled degradation. Second, we experimentally demonstrate SAL using an automated high-temperature characterization platform and Ga\textsubscript{2}O\textsubscript{3}-based diode sensors over \SIrange{350}{550}{\celsius} and H\textsubscript{2} concentrations of \SIrange{0}{800}{ppm}, showing that autonomous experimentation can generate curated, time-resolved IV datasets while preserving device integrity. Third, we develop a structured GP model for offline long-horizon forecasting of the device response at a target voltage, based on a Kohlrausch--Williams--Watts (KWW) degradation mean and residual covariance kernel, enabling forecasts over a 1,000-hour horizon from approximately 100 hours of training data.

\section{Active learning under time uncertainty and safety constrains}

When subjected to extreme environmental conditions—such as elevated temperatures and high H\textsubscript{2} concentrations—Ga\textsubscript{2}O\textsubscript{3}-based devices can undergo irreversible physicochemical changes, ultimately degrading their performance or leading to catastrophic failure. In this setting, classical active learning strategies that prioritize sampling solely based on model uncertainty reduction become hazardous, as they lack any mechanism to avoid unsafe operating regimes.

To enable fully autonomous characterization, the sampling policy must explicitly account for device safety. In practice, this requires defining a device-specific safety metric and constraining exploration to regions of the (temperature, H\textsubscript{2}) search space that the model predicts to be safe with high confidence. By enforcing such constraints, the system avoids irreversible device degradation while still efficiently guiding the experiment toward informative measurements.

A further complication arises from the temporal structure of the measurements themselves. The time required for each device to stabilize after a change in temperature or gas concentration is typically unknown a priori and varies significantly across operating conditions. This introduces \emph{time uncertainty} into the sampling process: we do not know when a measurement will be complete, nor when the next safe query can be issued. Traditional Bayesian optimization methods assume static evaluations with known or fixed durations, and therefore do not naturally handle this dynamic, asynchronous setting.

Taken together, these challenges motivate the use of a modified safe Bayesian optimization (SafeBO) framework—one adapted not only to enforce probabilistic safety constraints but also to operate under uncertain and condition-dependent measurement times. This Safe Active Learning (SAL) approach enables autonomous decision-making that remains both informative and safe throughout long-duration device characterization campaigns.

Bayesian optimization (BO) provides a framework for optimizing unknown functions using surrogate models, typically Gaussian processes (GPs), and acquisition functions that balance exploration and exploitation. However, in safety-critical settings, naive BO may query unsafe regions, motivating the development of Safe Bayesian Optimization (SafeBO).

SafeBO methods enforce that all evaluated points lie within a \emph{safe set}, defined as the region where one or more constraints are satisfied with high probability. These constraints are typically learned online using GP models and confidence bounds. A canonical formulation considers an unknown function $f(\mathbf{x})$ subject to a safety threshold $h$, requiring $f(\mathbf{x}_t) \ge h$ for all queried points.

The SafeOPT algorithm~\cite{sui2015a} starts from an initial safe seed set, and uses GP confidence intervals to iteratively expand the safe region while selecting informative points that either improve the objective or enlarge the safe set, ensuring safety at every step and convergence to the optimal safely reachable solution. STAGEOPT~\cite{sui2018a} extends this idea by separating safe region expansion and objective optimization into two distinct phases, simplifying the exploration--exploitation trade-off.

Subsequent work has generalized SafeBO in several directions. Extensions include handling multiple and contextual constraints~\cite{berkenkamp2023a} and improving theoretical guarantees through refined GP confidence bounds and Lipschitz-based formulations~\cite{fiedler2024a}. SafeBO has also been extended to more complex scenarios. Methods such as GoSafeOpt~\cite{sukhija2023a} enable exploration of disconnected safe regions in dynamical systems, while TVSAFEOPT~\cite{li2024a} introduces spatio-temporal modeling to track time-varying safe sets.

Closely related is the literature on \emph{safe active learning}, which focuses on learning accurate models under safety constraints rather than direct optimization. GP-based approaches design informative experiments while ensuring safety through predictive uncertainty or auxiliary models~\cite{zimmer2024b,lange-hegermann2025}, and can be extended to explore regions that satisfy performance constraints rather than a single optimum~\cite{malkomes2021a}. SafeBO has also been extended to high-dimensional and control-oriented settings, including embedding-based methods for large parameter spaces, control-theoretic safety enforcement along trajectories, and application-specific formulations with domain-defined safety metrics~\cite{wei2024a,vinter-hviid2024a,cole2024a}.

\begin{figure}[!t]
    \centering
    \includegraphics[width=\columnwidth]{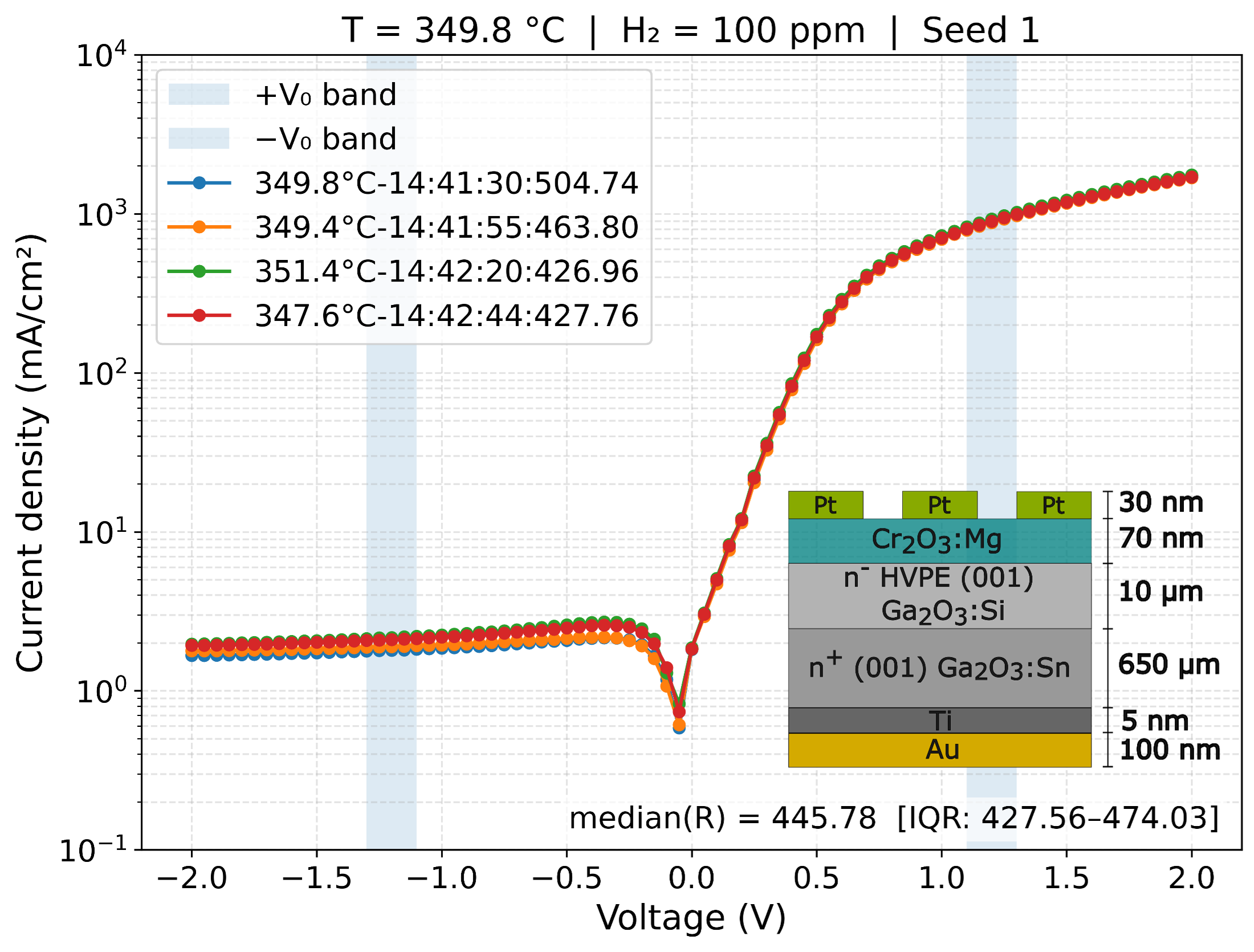}
    \caption{Representative set of four IV sweeps acquired during the acquisition of the first seed point of a SAL run at \(T=395.1~^\circ\mathrm{C}\) and \(H_2=250~\mathrm{ppm}\), illustrating the intra-band rectification calculation used online by the algorithm. In this case, the target bias was \(V_0 = 1.2~\mathrm{V}\), and the shaded regions indicate the fixed voltage bands of \(\pm 0.1~\mathrm{V}\) around the target bias magnitudes \(\pm V_0\). Within these bands, forward- and reverse-bias points with the closest matching \(|V|\) values are paired to compute intra-band rectification ratios, from which the sweep-level rectification is obtained as the median of the valid pairwise ratios. The figure also reports the median rectification across the four sweeps and the corresponding interquartile range (IQR), highlighting the robustness of the aggregation against sweep-to-sweep variability during SAL operation.}
    \label{fig:iv_sample}
\end{figure}

We introduce a variant of the previously described SafeBO-style algorithms that operates on the time–temperature–concentration rectification surface
\begin{equation}
  R = R(t, T, G),
\end{equation}
where \(t\) is the elapsed campaign time, \(T\) is the device temperature, and \(G\) is the \(\mathrm{H}_2\) concentration. The device under test is a rectifying \(\mathrm{Ga_2O_3}\)-based diode. For this class of devices, the rectification ratio at a given bias reflects the quality of the junction barrier and the extent of reverse leakage: a sustained drop in rectification is typically associated with barrier degradation, activation of additional leakage paths (for example, defect-assisted conduction), or partial device failure. Conversely, maintaining a large forward-to-reverse current ratio indicates that the depletion region, barrier height, and leakage mechanisms remain in a regime consistent with normal operation. We therefore use rectification as a scalar safety observable and enforce a minimum rectification threshold as a proxy for “safe operation’’ of the device under combined thermal and hydrogen exposure, over time. Fig.~\ref{fig:iv_sample} shows IV samples from a real Pt/Cr\textsubscript{2}O\textsubscript{3}:Mg/$\beta$-Ga\textsubscript{2}O\textsubscript{3}, discussed in section~\ref{sec:validation} and the corresponding intra-band calculation of the rectification factor for a set of sequential IV sweeps, as discussed in the appendix~\ref{A1}.

The Safe Active Learning (SAL) algorithm couples this rectification-based safety notion with a Gaussian-process (GP) surrogate over \((t,T,G)\), an adaptive completion-time window that accounts for uncertain experiment durations, a trust region in \((T,G)\) anchored to previously observed safe conditions, and a two-phase sampling schedule that starts from a conservative rectification threshold and then gradually relaxes it. All GP modeling and acquisition logic are implemented directly in BoTorch, using a custom SingleTaskGP with an additive kernel. A complete description of the SAL framework, including its mathematical formulation, operational phases, safety constraints, implementation parameters, is provided in Appendix~\ref{A1}. SAL
s pseudo-code is shown in Fig.\ref{fig:pseudocode}.

\begin{figure}[!t]
\centering
\begin{tcolorbox}
\small
\textbf{Safe Active Learning (SAL)}

\vspace{4pt}
\textbf{1. Seed initialization}
\begin{enumerate}
\item Run experiments at seed $(T,G)$ points.
\item Compute rectification $R$ from IV data; ban invalid conditions.
\item Initialize dataset $\mathcal{D}$ with $(t,T,G,R)$.
\end{enumerate}

\vspace{4pt}
\textbf{2. Phase 1: Safe exploration at fixed threshold $h$ (for $i=1$ to $N_1$)}
\begin{enumerate}
\item Fit GP model on $\log R(t,T,G)$.
\item Construct adaptive completion-time window from recent durations.
\item Compute time-window lower bound $L_{\mathrm{win}}(T,G)$.
\item Build safe set $\mathcal{S}_{\mathrm{safe}}=\{(T,G):L_{\mathrm{win}}\ge h\}$.
\item Intersect with trust region around measured-safe points.
\item If empty: relax $\beta$; if still empty, invoke rescue.
\item Compute acquisition
\item Select $(T,G)$ maximizing weighted acquisition inside safe set.
\item Execute experiment, update dataset and durations, ban invalid points.
\end{enumerate}

\vspace{4pt}
\textbf{3. Phase 1 Rescue (if safe set collapses)}
\begin{enumerate}
\item Re-measure most recent safe condition.
\item Classify outcome as modeling artifact, boundary behavior, or failure.
\item Resume Phase 1 (remaining budget), transition to Phase 2, or terminate.
\end{enumerate}

\vspace{4pt}
\textbf{4. Phase 2: Threshold relaxation (for $j=1$ to $N_2$)}
\begin{enumerate}
\item Update target $\tau_k$ via exponential decay.
\item If $\tau_k>1+\epsilon$:
\begin{itemize}
\item Repeat Phase~1 logic using threshold $\tau_k$.
\item If safe set empty: switch to trust-region uncertainty fallback.
\end{itemize}
\item If $\tau_k\approx1$: drop safety gating and maximize uncertainty globally.
\item Execute experiment and update model.
\end{enumerate}

\end{tcolorbox}
\caption{High-level pseudocode of the Safe Active Learning (SAL) algorithm. Phase~1 and Phase~2 operate under fixed iteration budgets $N_1$ and $N_2$, respectively.}
\label{fig:pseudocode}
\end{figure}

\section{Evaluation of the SAL algorithm on simulated data}

To assess the effectiveness of the SAL algorithm in modeling the rectification factor over the $(t,T,G)$ space, we first evaluate its behavior on simulated data prior to experimental validation. In this simulation mode, laboratory calls are replaced by evaluations of an analytic rectification model. A full description of the simulation setup is provided in Appendix~\ref{A2}, and only the results are presented here.

\begin{figure*}[!t]
    \centering
    \includegraphics[width=\textwidth]{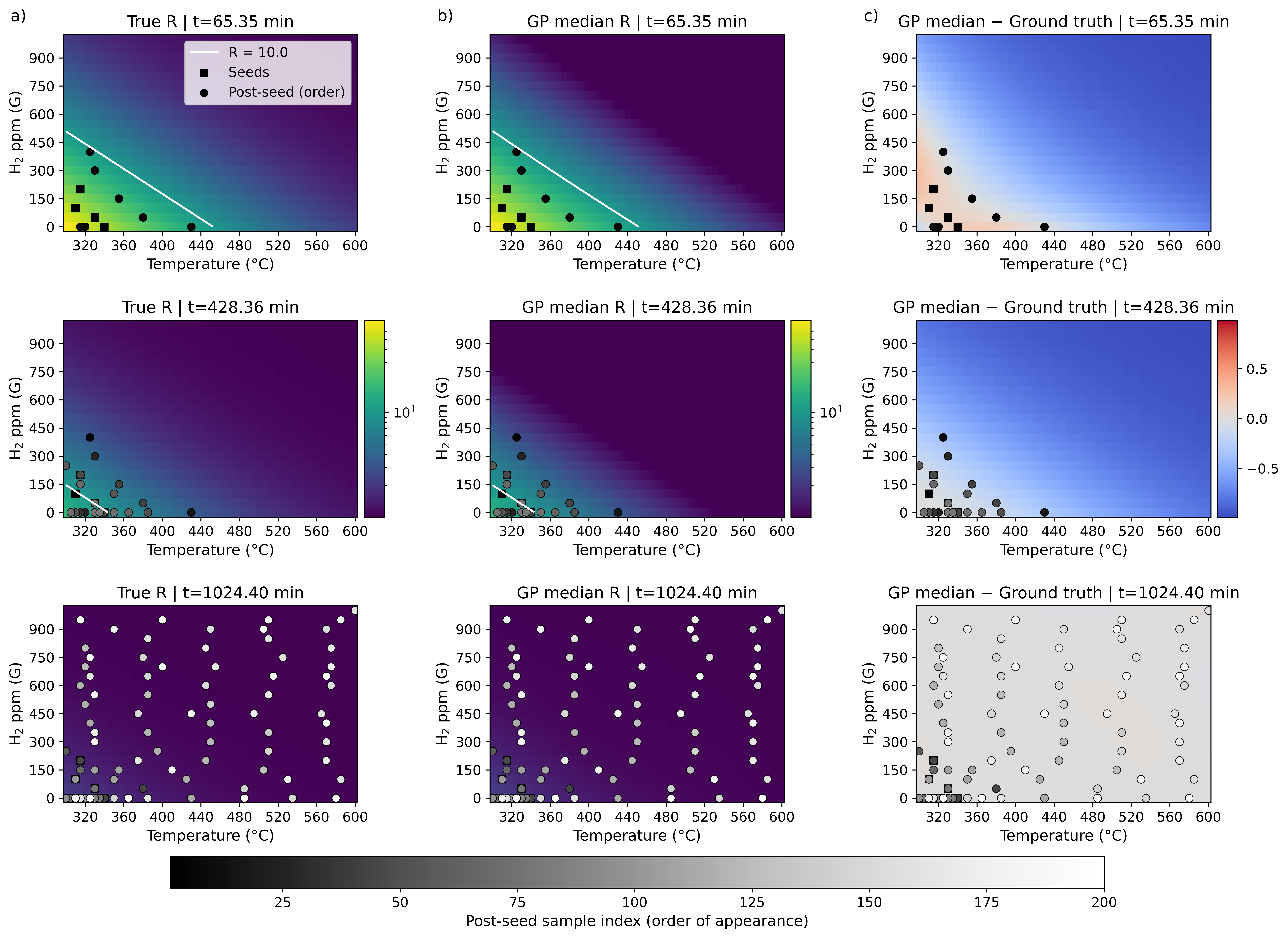}
    \caption{Residual analysis across selected SAL iterations. Each row corresponds to a snapshot at a specific global iteration, using the posterior mean surface $GP(T,G)$ and associated time $t_{\mathrm{now}}$ saved during the SAL run. Columns show (a) the ground-truth rectification surface $R_{\mathrm{true}}(T,G)$, (b) the GP posterior median surface $GP(T,G)$ at the same time, and (c) the residual surface $GP(T,G) - R_{\mathrm{true}}(T,G)$. White contours indicate the safety threshold $R = 10$. Overlaid markers denote sampled $(T,G)$ conditions up to the corresponding iteration, with squares representing seed points and circles indicating post-seed samples ordered by sampling step.} 
    \label{fig:simulations_snapshots_noiseless}
\end{figure*}

Fig.~\ref{fig:simulations_snapshots_noiseless} presents representative SAL snapshots at three distinct time instants of the campaign. The left column shows the ground-truth rectification surface $R_{\mathrm{true}}(T,G)$, the middle column the GP posterior median $GP(T,G)$, and the right column the residual $GP(T,G)-R_{\mathrm{true}}(T,G)$. Since the GP is trained in log space, visualizations are reported using the median of $R$ and the interval $\exp(\mu_{\log} \pm 2\sigma_{\log})$, as detailed in the ESI.

In phase 1 (top and middle rows of Fig. \ref{fig:simulations_snapshots_noiseless}), sampling is confined strictly within the initial safe region ($R \geq 10$) and as the campaign time elapses, the safe region shrinks (as defined in section \ref{ground_truth}). The GP rapidly captures the global monotonic structure of the surface despite sparse data, with residuals remaining small and spatially smooth.

During Phase 2, as $\tau$ decreases according to Eq.~\ref{tau_schedule}, the admissible region expands in a controlled manner. Sampling gradually approaches higher $(T,G)$ conditions while remaining within the dynamically updated safety boundary. The GP continues to track the evolving surface accurately, with no evidence of boundary misclassification.

In the late stage (bottom row of Fig. \ref{fig:simulations_snapshots_noiseless}), after the transition to pure uncertainty exploration, sampling spans the entire $(T,G)$ domain. The residual surface becomes uniformly small, indicating global convergence of the surrogate model.

Overall, these results demonstrate that SAL maintains strict safety compliance while progressively expanding the search region and achieving accurate global reconstruction of the rectification surface. ESI \textit{simulation movie} shows the sampling evolution of SAL, showing how the GP mean, standard deviation, LCB and acquisition surface evolve as the algorithm progresses from phase 1, to pure exploration in phase 2 (pure uncertainty mode).

To assess whether SAL recovered the true model governing the rectification factor R, Figs. S2 and S3 in the ESI compare time slices at several (T,G) conditions between the true model and the GP surrogate learned from SAL. In densely sampled regions (Fig. S2), the GP median closely overlaps the true trajectories and the measured R data, yielding uniformly low errors and tight uncertainty bands. In data-scarce regions (Fig. S3), the behavior is more nuanced than a monotonic degradation with reduced sampling. As expected, when a (T,G) condition lies outside the region effectively constrained by SAL observations, the GP exhibits broader uncertainty and can incur larger RMSE due to extrapolation. 

However, several sparsely (or even un-)sampled conditions are still predicted with very small RMSE, indicating that the GP can generalize accurately when the target condition remains strongly constrained by learned correlations and smooth trends across (T,G) and time. Overall, these results show that SAL not only fits the model well where it samples densely, but also captures the underlying functional structure well enough to provide accurate predictions in parts of the domain with limited direct data support—while appropriately signaling uncertainty where extrapolation is required.

We note that the results discussed above correspond to a noiseless scenario, i.e., the simulated measurements were not corrupted by noise. To assess the robustness of SAL under more realistic conditions, we also considered a simulation with increased measurement noise (10\% relative noise with a 0.5 absolute floor). Under this noisier regime, SAL still captures the main structure of the rectification model, although the impact of noise becomes clearly visible in both the learned GP surfaces and the time-slice comparisons (Figs. S4-S6 of the ESI). In regions with dense sampling (Fig. S4), the GP median remains close to the true trajectories and follows the measured data well, albeit with wider uncertainty bands and slightly larger RMSE compared to the noiseless case. In sparsely sampled regions (Fig. S5), the GP relies more heavily on extrapolation, which can lead to larger deviations from the true model and higher RMSE values for some (T,G) conditions. 

Figure S6 shows snapshots of the GP-predicted rectification surfaces at different times during the campaign, illustrating how the surrogate model progressively reconstructs the underlying rectification landscape despite the presence of measurement noise. The GP predictions still recover the correct qualitative trends across time, temperature, and gas concentration, while the model uncertainty appropriately expands in regions where the data provide weaker constraints. Overall, these results demonstrate that SAL remains capable of learning the underlying model structure even under significantly noisier measurements, while properly reflecting reduced confidence in poorly constrained regions of the search domain.

\section{Instrumentation and system orchestration}

\begin{figure}[t]
    \centering
    \includegraphics[width=\columnwidth]{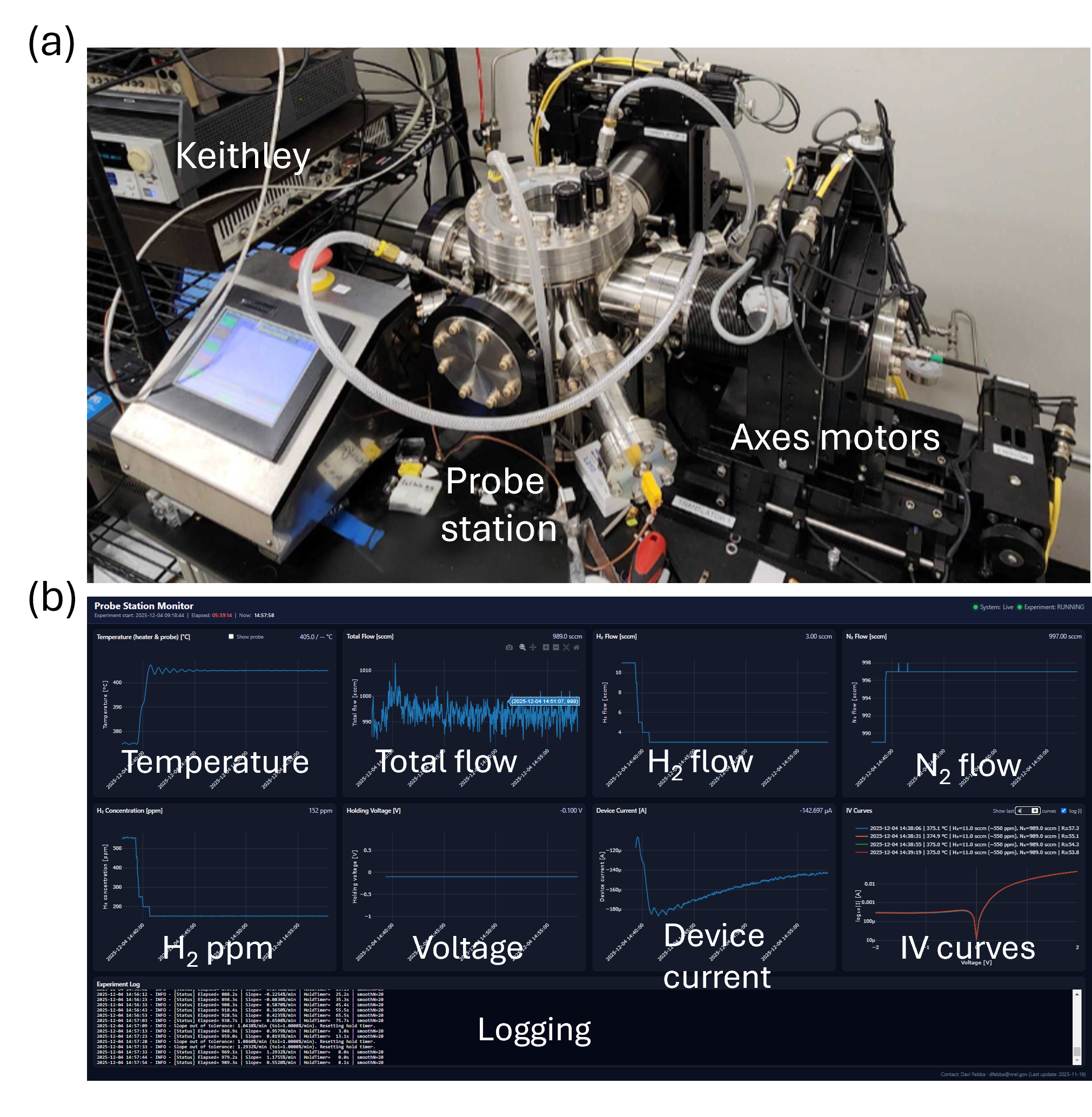}
    \caption{a) The high-temperature probe station used in this work, with key components, and b) the web application developed to provide real-time visualization of system telemetry and experiment status}
    \label{fig:probe_station}
\end{figure}

Device characterization was carried out using an automated high-temperature probe station, shown in Fig.~\ref{fig:probe_station}a. The station employs a resistive heating element to control the temperature of a stage inside a vacuum chamber, on which the device is mounted. Two Alicat mass flow controllers (MFCs) regulate the flow of nitrogen and forming gas containing \(5\,\%\) \(\mathrm{H_2}\), and an Alicat flow meter at the exhaust measures the total outlet flow from the chamber.

Two probe arms contact the device: one on the top \(1~\mathrm{mm}\)-diameter electrode, and the other on an \(\mathrm{Al_2O_3}\) substrate that supports the device back contact. The device is secured to the \(\mathrm{Al_2O_3}\) fixture using a mechanical clamp, with Ag paint applied between the fixture and the back contact to reduce the contact resistance. A Keithley 2400 source-measure unit (SMU), operated in two-wire mode with auto-ranging enabled, is used to execute IV sweeps and to record the device current as a function of time.

All individual components—probe station, MFCs, flow meter, and SMU—are controlled by a custom system orchestrator. This orchestrator coordinates and executes all steps of the active-learning experiments driven by the SAL algorithm, including asynchronous equilibrium checks, continuous telemetry acquisition, IV sweep execution, error handling, and real-time event logging. An application programming interface (API) was developed for the orchestrator to facilitate user customization of experimental workflows and the addition or removal of functionality. Moreover, a web application was developed to provide real-time visualization of telemetry data and experiment progress, as shown in Fig.~\ref{fig:probe_station}b. More details of this web application can be seen in Fig. S1 of the ESI.

\begin{figure*}[!t]
    \centering
    \includegraphics[width=\textwidth]{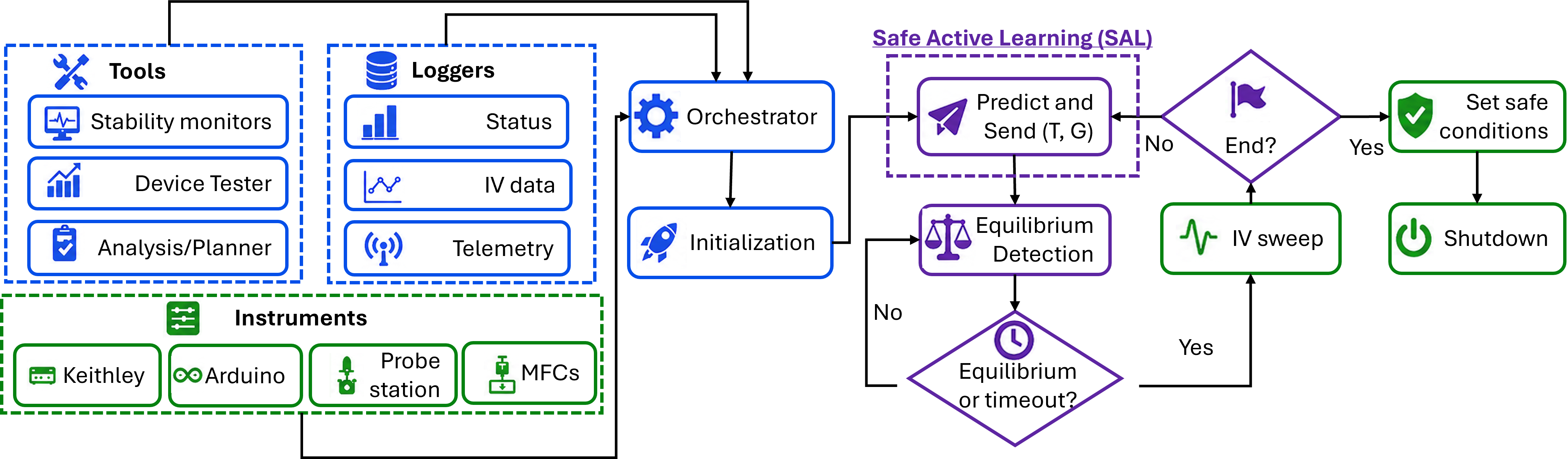}
    \caption{High-level system architecture and workflow illustrating orchestrator initialization and the automated IV measurement sequence. After SAL predicts new operating conditions \((T,G)\), these settings are applied to the probe station, and equilibrium detectors assess system stabilization (or timeout) before triggering an IV sweep.}
    \label{fig:probe_workflow}
\end{figure*}

Fig. \ref{fig:probe_workflow} shows the different modules required by the system orchestrator to perform communications with several instruments, logging and measurements, as well as when IV sweep measurements are taken based on equilibrium detection after a new set of (T,G) is sent to the probe station.

IV sweeps to measure \(R\) are executed only after a sequence of equilibrium-detection routines has completed. Each time a new set of \((T,G)\) conditions is sent to the instruments, two detectors run in parallel. First, a temperature-equilibrium detector monitors the sample temperature at 1~Hz. Using the most recent \SI{60}{\second} of data, equilibrium is declared when the rate of change of the temperature falls below \(1\%\)/min and this condition is maintained for five consecutive checks.

Second, a gas-flow detector monitors the inlet \(\mathrm{N_2}\) and forming-gas flows, as well as the exhaust flow. All flow readings, expressed in standard cubic centimeters per minute (sccm), must lie within user-defined tolerances of their respective setpoints for the chamber to be considered flow-stable. In addition, the exhaust flow must remain within a tolerance band around the total inlet flow (sum of \(\mathrm{N_2}\) and forming gas).

Once both environmental conditions are stable, a third detector tracks the device current at a user-defined holding voltage. Using the most recent \SI{60}{\second} of current data, smoothed over 20 points, equilibrium is declared when the rate of change of the current falls below \(1\%\)/min and this condition is sustained for \SI{2}{\minute}. Only after this device-current equilibrium detector returns successful does SAL proceed with the execution of IV sweeps.

These equilibrium mechanisms ensure that each rectification measurement is taken under well-defined thermal, gas-flow, and electrical conditions, but they also make the effective completion time of an experiment strongly condition-dependent. The time required to reach equilibrium can vary significantly with the distance from the previous operating point, the thermal inertia of the system, and the device state, so SAL only knows \emph{when} a measurement will complete after the fact. From the algorithm’s perspective, the completion time associated with a planned \((T,G)\) is thus a random quantity, and rectification must be modeled as a function of time as well as \((T,G)\). This motivates the introduction of an adaptive completion-time window, within which SAL reasons about safety, as previously discussed.

\begin{figure}[!t]
    \centering
    \includegraphics[width=\columnwidth]{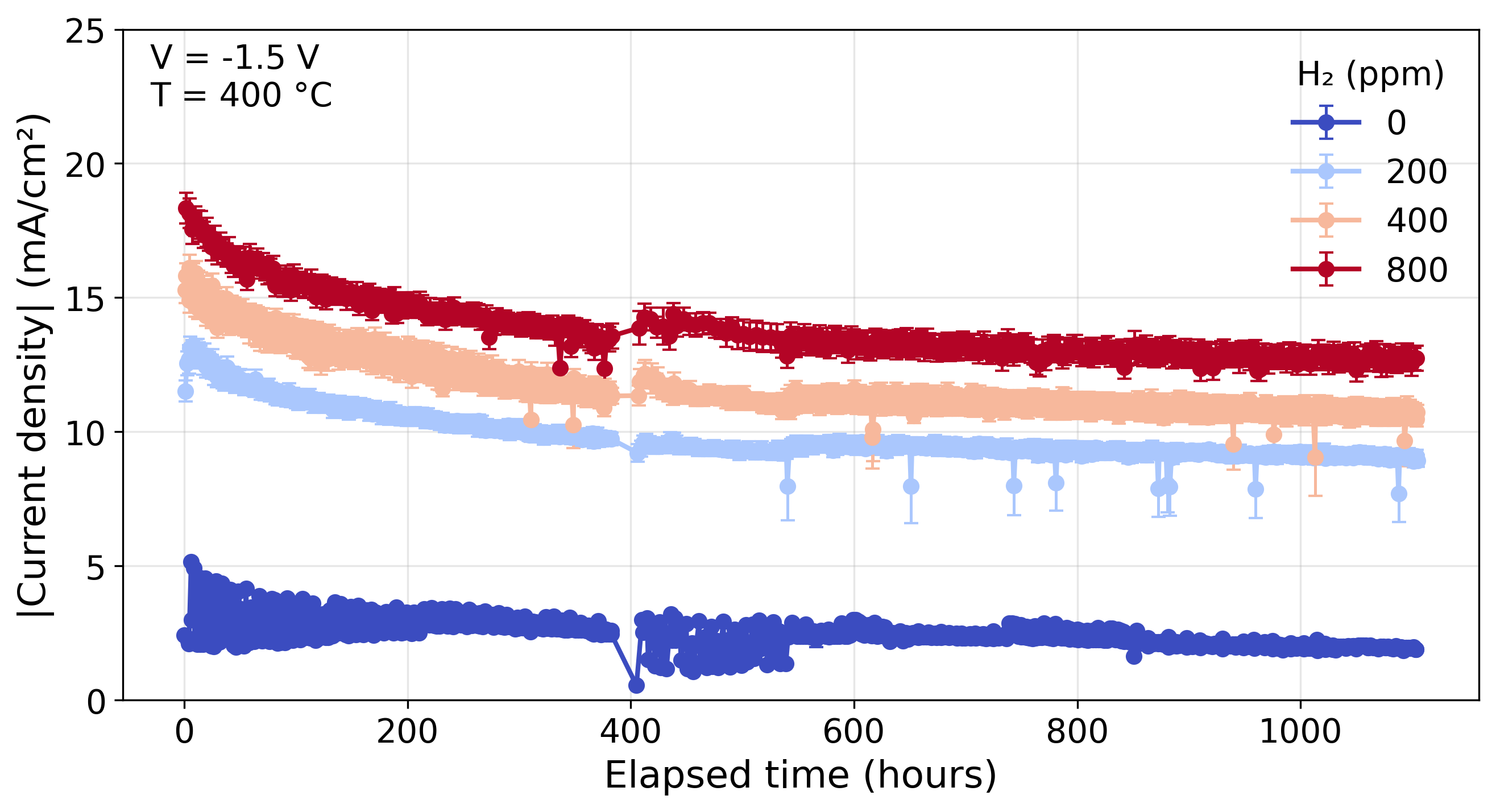}
    \caption{Validation dataset collected from a device fabricated in a different batch as the SAL-tested device. The current density at \(-1.5~\mathrm{V}\) is shown as a function of elapsed time at \(T = 400~^\circ\mathrm{C}\) for hydrogen concentrations of 0, 200, 400, and 800~ppm. Each data point represents the mean over a block of four sequential IV sweeps, while error bars represent the standard error of the mean (SEM). The data reveal a systematic dependence of the device response on both hydrogen concentration and time, and are used to validate long-horizon Gaussian process (GP) predictions at fixed \((T, G)\) conditions.}
    \label{fig:validation_dataset}
\end{figure}

\section{Device fabrication and Validation dataset}
\label{sec:validation}

We choose to test the SAL algorithm and autonomous measurements on the case of Pt/Cr\textsubscript{2}O\textsubscript{3}:Mg/$\beta$-Ga\textsubscript{2}O\textsubscript{3} diode sensors of hydrogen and temperature, because the Cr\textsubscript{2}O\textsubscript{3}/$\beta$-Ga\textsubscript{2}O\textsubscript{3} heterojunction has been shown to be more stable\cite{saha_cr2o3-ga2o3_2026} than then more commonly studied NiO/$\beta$-Ga\textsubscript{2}O\textsubscript{3} heterojunction\cite{egbo_niga2o4_2024,saha_cr2o3-ga2o3_2026} which rapidly degrades at high temperature and reduces in hydrogen environment. A Pt/Cr\textsubscript{2}O\textsubscript{3}:Mg/$\beta$-Ga\textsubscript{2}O\textsubscript{3} vertical heterojunction diode (inset of Fig.~\ref{fig:iv_sample}) was fabricated using a \SI{10}{\micro\meter} lightly Si-doped (\SI{3e16}{\per\cubic\centi\meter}) n-type $\beta$-Ga\textsubscript{2}O\textsubscript{3} drift layer grown on a conductive bulk (001) $\beta$-Ga\textsubscript{2}O\textsubscript{3}:Sn substrate (NCT).\cite{febba2025a} Photoresist was removed from the as-delivered substrates via an organic wash followed by a piranha rinse. Following photoresist removal, the substrates were further cleaned via UV ozone exposure for \SI{10}{\minute} at room temperature. Large-area back Ohmic contacts were then deposited via a bi-metal deposition of \SI{5}{\nano\meter} Ti/\SI{100}{\nano\meter} Au\cite{callahan2023} using a Temescal FC2000 evaporation system under high vacuum conditions (\SI{3e-6}{\torr}) without venting between layers. Contacts were annealed in flowing nitrogen at \SI{100}{\celsius} for \SI{1}{\minute} as a preliminary cleaning step, followed by \SI{550}{\celsius} for \SI{90}{\second}. The p-type Cr\textsubscript{2}O\textsubscript{3}:Mg layer was deposited via pulsed laser deposition (PLD). During PLD growth, a 1 inch ceramic target of 8~at.\% Mg-doped Cr\textsubscript{2}O\textsubscript{3} was ablated using a pulsed KrF excimer laser ($\lambda = \SI{248}{\nano\meter}$) at a frequency of \SI{10}{\hertz} and energy of \SI{300}{\milli\joule}\cite{callahan2024a}. Growth was performed at a substrate temperature of \SI{600}{\celsius} and \ce{O2} partial pressure of \SI{20}{\milli\torr}, resulting in a thickness of \SI{70}{\nano\meter} as determined by X-ray reflectivity. The \SI{30}{\nano\meter}-thick, \SI{1}{\milli\meter}-diameter hydrogen sensing Pt pads were deposited via shadow mask using the same e-beam evaporation system as previously described. To improve metal adhesion to the Cr\textsubscript{2}O\textsubscript{3} layer, the Pt deposition rate was set to \SI{0.5}{\angstrom\per\second} for the first \SI{10}{\nano\meter}, followed by \SI{1}{\angstrom\per\second} for the remaining \SI{20}{\nano\meter}. A representative IV curve obtained from one of these devices is show in Fig.~\ref{fig:iv_sample}.

To formulate and stress-test the long-horizon forecasting model used for offline response analysis, we collected an auxiliary validation dataset from a device belonging to a different batch as the one subjected to SAL. This validation campaign was not designed to reproduce the full SAL measurement history. Instead, it was designed as a controlled benchmark for long-time modeling of the device current at a fixed target voltage, under a simplified exposure protocol. In particular, the temperature was held fixed throughout the experiment, so the dataset does not contain temperature cycles or repeated excursions across a broad \((T,G)\) space, unlike the dataset that SAL will generate during autonomous characterization.

The same procedure was repeated at increasing H\textsubscript{2} concentrations of 200, 400, and \SI{800}{ppm}. Subsequently, the H\textsubscript{2} concentration was decreased in reverse order down to zero flow. This entire sequence was repeated until the total exposure time exceeded 1,000 hours. Fig.~\ref{fig:validation_dataset} shows the resulting validation dataset at a target voltage of \SI{-1.5}{\volt}.

A few features of Fig.~\ref{fig:validation_dataset} are worth noting. First, the current exhibits a clear time dependence at all four gas concentrations, with an initially faster evolution followed by a slower long-time drift, indicating that elapsed exposure time is an important
variable for modeling the device response. Second, the signal remains ordered with hydrogen concentration over most of the experiment, but the separation between adjacent concentrations is modest. In other words,
under these conditions the device is not highly sensitive to H\textsubscript{2}: the effect of gas concentration is measurable and systematic, but substantially weaker than would be expected for a
strongly selective hydrogen sensor. Instead, the response is dominated by a common time-dependent degradation trend, with hydrogen concentration acting primarily as a secondary perturbation that shifts the current level and only modestly modifies the degradation trajectory.

This structure makes the validation dataset particularly useful for kernel formulation. Because temperature is fixed, the data isolate the dominant long-time evolution of the device response and expose the combination of (i) a monotone degradation trend, (ii) partial flattening at long times, and (iii) comparatively weak but systematic gas-dependent offsets. These features motivate the temporal form of the
structured long-horizon model introduced in Section~\ref{sec:offline_kernel}. Importantly, however, the resulting model is not restricted to fixed-temperature experiments: its mean and residual terms are defined over the full \((t,T,G)\) space, so it can be fit to SAL-generated datasets containing multiple temperature levels and temperature cycles.

\begin{figure}[!t]
    \centering
    \includegraphics[width=\columnwidth]{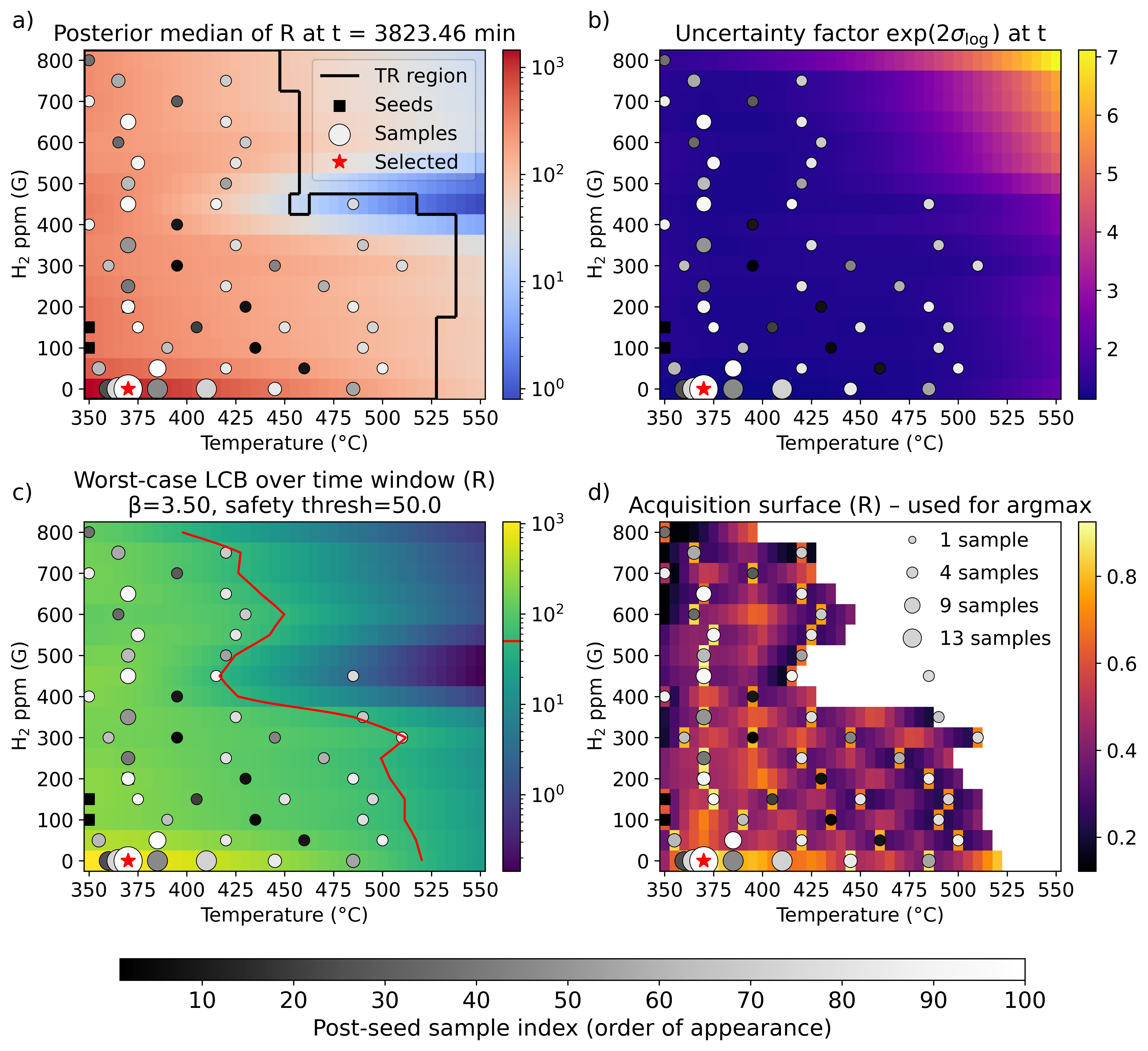}
    \caption{Representative late-Phase-1 SAL snapshot at iteration 100. (a): posterior median rectification surface at the current time instant, with all sampled conditions overlaid; marker size indicates repeated visits to the same \((T,G)\) condition, black squares denote seed points, open circles denote post-seed measurements colored by order of acquisition, the white boundary marks the trust region, and the red star indicates the next selected condition. (b): multiplicative uncertainty factor \(\exp(2\sigma_{\log})\), visualizing posterior uncertainty in the log-normal rectification model. (c): worst-case lower-confidence-bound rectification surface evaluated over the completion-time window used by SAL; the red contour marks where this conservative surface reaches the safety threshold (\(R=50\)), defining the current safety boundary. (d): acquisition surface used to select the next experiment; white regions are excluded from selection after safety and admissibility gating, and the red star marks the selected maximizer.}
    \label{fig:sal_iter_100}
\end{figure}

\section{SAL-driven characterization of a temperature- and H\textsubscript{2}-sensing Pt/Cr\textsubscript{2}O\textsubscript{3}:Mg/$\beta$-Ga\textsubscript{2}O\textsubscript{3} device}

The parameters used for the autonomous SAL campaign on the Pt/Cr\textsubscript{2}O\textsubscript{3}:Mg/$\beta$-Ga\textsubscript{2}O\textsubscript{3} device are listed in Table S1\dag. Figure~\ref{fig:sal_iter_100} shows a representative snapshot of the full SAL state at the end of phase 1, and a movie showing the evolution of the SAL surfaces throughout the entire run is provided in the ESI.

The figure summarizes how SAL combines prediction, uncertainty quantification, safety filtering, and acquisition-driven experiment selection in the \((T,G)\) domain. The top-left panel shows the posterior median rectification surface together with all conditions visited up to that point. Marker size reflects repeated visits to the same \((T,G)\) condition, indicating that SAL does not simply sweep the grid uniformly, but instead selectively revisits conditions when additional information is valuable. The white boundary marks the trust region, i.e., the subset of \((T,G)\) space currently reachable from previously verified safe conditions. This highlights an important feature of SAL: exploration remains anchored to empirically supported regions rather than expanding freely across the full search domain.

The top-right panel shows the multiplicative uncertainty factor used to visualize posterior spread in rectification. Because the GP is trained in log-rectification, the back-transformed posterior for rectification is approximately log-normal rather than Gaussian. For this reason, the central prediction is shown as the posterior median rather than the arithmetic mean, and uncertainty is reported as a multiplicative factor rather than as an additive standard deviation. This representation is more appropriate for a strictly positive quantity that spans orders of magnitude and is directly consistent with the internal safety logic of SAL.

The bottom-left panel shows the worst-case lower-confidence-bound rectification surface over the completion-time window considered by SAL at that iteration. This is the key safety surface used online: it is a conservative summary over plausible experiment-completion times, rather than a nominal prediction at a single future time. The red contour marks where this lower-confidence-bound surface is equal to the safety threshold used in the run. Conditions on the safe side of this contour remain admissible, whereas conditions beyond it are excluded because their conservative predicted rectification falls below the required threshold. The nontrivial shape of this boundary illustrates that safety is strongly condition dependent, with the accessible region shrinking in harsher parts of the temperature--hydrogen space.

The bottom-right panel shows the acquisition surface actually used to select the next experiment. Importantly, this surface is not determined by uncertainty alone. In Phase~1, SAL combines a time-aggregated uncertainty term with an exploration term that favors coverage of under-sampled regions, a revisit term that can return the algorithm to previously sampled conditions after sufficient time has elapsed, and a penalty that discourages persistent edge-hugging near the domain boundaries. This composite score is then hard-gated by the admissible set, so only conditions that satisfy the current safety criterion and trust-region constraint remain eligible; permanently banned points are excluded. White regions are therefore excluded from selection, while the red star marks the condition chosen by maximizing the resulting acquisition surface.

During the SAL campaign, only one unsafe measurement was observed in Phase~1. This event was associated with a spurious batch of IV sweeps (Fig.~S7\dag) that showed unusually large variability, leading to an unreliable rectification measurement. No \((T,G)\) conditions were banned during Phase~1. In Phase~2, 24 of 100 iterations produced rectification factors below the fixed Phase~1 threshold, as expected because the safety threshold is intentionally relaxed during this stage. Of these, 9 were also unsafe relative to the iteration-specific Phase~2 threshold. Because the device progressively degrades over time, this behavior is expected. Taken together, these results indicate that SAL can explore the domain conservatively and with low probability of safety violations in Phase~1, while intentionally pushing the device toward riskier operating conditions in Phase~2, as intended.

\section{Offline modeling and long-horizon forecasting with the KWW-based GP}
\label{sec:offline_kernel}

The kernel used by SAL (Eq.~\eqref{SAL_kernel}) is well suited to the online decision-making objective of the active-learning campaign. Its additive RBF + linear form provides smooth local interpolation while
retaining enough global structure to avoid posterior collapse between widely separated training points. Moreover, during the campaign SAL only evaluates the surrogate at times close to the currently observed
completion times, so predictive behavior far beyond the sampled time range is not required. For this reason, the SAL kernel is appropriate for local safe exploration, but not for the offline forecasting task
considered here. In particular, the SAL model uses a constant prior mean, so its posterior tends to revert toward an uninformative baseline once the prediction time moves far beyond the observed interval.
Likewise, the additive linear term can impose an unbounded temporal drift, which is undesirable for degradation trajectories that are expected to flatten at long times. More generally, a purely local kernel
of this type does not encode any explicit prior expectation of monotone, saturating degradation.

The response-modeling analysis pursued here has a different objective. Rather than selecting the next safe experiment, it aims to model the device response at a target voltage, \(|I(V_{\mathrm{target}})|\), over long time horizons and ultimately to support offline forecasting on the SAL-generated dataset across the full \((t,T,G)\) domain. The auxiliary validation dataset of Section~\ref{sec:validation} is used here to inform the construction of that forecasting model because it isolates a simpler but highly
informative regime: fixed temperature, stepped gas concentration, and long-duration exposure without temperature cycling. Under these conditions, the current traces exhibit a shared monotone time dependence
with an initially faster decay followed by gradual flattening at long times, while the effect of H\textsubscript{2} concentration appears primarily as a modest, systematic shift between trajectories. This provides a clean basis for choosing the temporal form of the long-time degradation prior. Importantly, however, the resulting model is not restricted to that fixed-temperature regime. Both its structured mean and its residual GP depend explicitly on \(t\), \(T\), and \(G\), so the same model family can be trained on SAL-generated datasets containing multiple temperature levels and temperature cycles. 

Developing such a model required a formulation capable of long-term forecasting while retaining the observed temperature- and gas-dependent trends. Agentic LLM workflows provide a credible framework for combining literature-grounded reasoning with scientific computation and physical-model development.\cite{yao_operationalizing_2025} Following this approach, we used GitHub Copilot with Anthropic models to assist in exploring physics-aware GP formulations. The agents were given access to the SAL training data, the auxiliary validation dataset, and the original SAL kernel, and were tasked with identifying a formulation capable of long-term forecasting while capturing the observed trends in \((T,G)\). This AI-assisted model-engineering process supported the following representation of the response as a structured degradation trend plus a GP residual:

\begin{equation}
f(x) = \mu_{\mathrm{KWW}}(x) + g(x),
\label{eq:kww_plus_residual}
\end{equation}
where
\begin{equation}
g(\cdot) \sim \mathcal{GP}\!\left(0,\,k_{\mathrm{res}}(\cdot,\cdot)\right)
\label{eq:residual_gp}
\end{equation}
Equivalently,
\begin{equation}
f(\cdot) \sim
\mathcal{GP}\!\left(\mu_{\mathrm{KWW}}(\cdot),\,
k_{\mathrm{res}}(\cdot,\cdot)\right)
\label{eq:full_gp}
\end{equation}

The choice of \(\mu_{\mathrm{KWW}}\) was motivated by the qualitative form of the observed degradation curves and by the broader use of stretched exponentials to describe relaxation and aging in heterogeneous
solid-state systems.\cite{johnston2006,fery2005} A simple exponential decay assumes a single characteristic time scale, whereas the present device response evolves more gradually and tends to
flatten at long times, indicating a distribution of effective time constants rather than a single rate-limiting process. The Kohlrausch--Williams--Watts (KWW) function provides a compact phenomenological description of this behavior: it remains monotone, permits sub-exponential long-tailed decay, and approaches a finite long-time level rather than diverging. These properties make it a more appropriate prior for long-horizon forecasting than either constant mean reversion or an unbounded linear trend.

Accordingly, we define the prior mean in log-current space as a condition-dependent KWW degradation law,
\begin{equation}
\mu_{\mathrm{KWW}}(t,T,G)
=
B(T,G)
-
A(T,G)
\left[
1-\exp\!\left(-[\lambda(T,G)t]^{\beta}\right)
\right]
\label{eq:kww_mean}
\end{equation}
Here \(B(T,G)\) is the condition-dependent baseline, \(A(T,G)\) is the degradation amplitude, and \(\lambda(T,G)\) is the effective degradation rate. This decomposition is useful because these three quantities play distinct physical roles. The baseline \(B(T,G)\) captures the response expected at the start of a dwell under a given thermal and gas environment, before appreciable degradation has 
accumulated. The amplitude \(A(T,G)\) determines how far the response can drop from that baseline under prolonged exposure, thereby controlling the long-time degradation depth. The rate \(\lambda(T,G)\) controls how quickly the system approaches that degraded state. Introducing separate functions for baseline, amplitude, and rate therefore allows the model to distinguish between a condition that changes the initial current level, one that increases the eventual severity of degradation, and one that primarily accelerates the time scale of degradation. More details about the proposed kernel are presented in the appendix~\ref{A3}, and here we focus our attention on the long-horizon results obtained by this approach.

\begin{figure}[!t]
    \centering
    \includegraphics[width=\columnwidth]{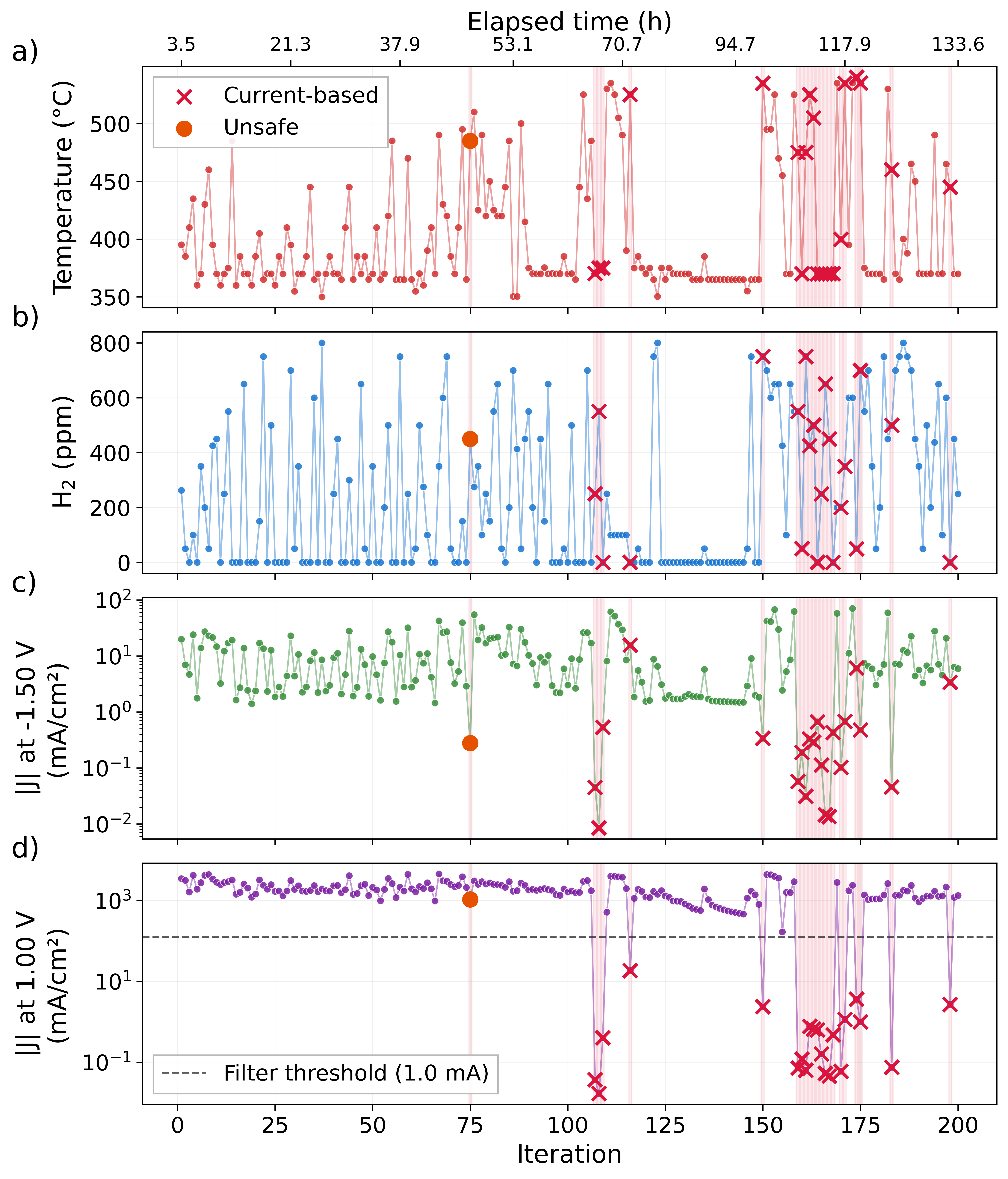}
    \caption{Timeline of SAL operating conditions and removed measurements during the autonomous campaign. (a) shows the selected temperature; (b) the selected H\textsubscript{2} concentration; (c) the measured current density at \SI{-1.5}{\volt}, and (d) the measured current density at \SI{1.0}{\volt} as a function of iteration; the top axis reports the corresponding elapsed time. Red crosses mark measurements rejected by the current-based quality filter, while orange circles denote measurements classified as unsafe. The dashed horizontal line in the bottom panel indicates the \SI{1.0}{\milli\ampere\per\centi\meter\squared} filter threshold used to identify severely degraded forward-bias sweeps. The figure shows that most discarded measurements occur late in the campaign, primarily during Phase~2, when SAL intentionally explores increasingly aggressive operating conditions.}
    \label{fig:bad_data}
\end{figure}

Unlike the auxiliary validation campaign of Section~\ref{sec:validation}, the SAL-generated dataset contains repeated temperature and gas transitions throughout the experiment and therefore reflects a substantially richer exposure history. At the same time, the two-phase structure of SAL means that, during Phase~2, the algorithm intentionally drives the device toward progressively riskier operating conditions. For this reason, all measurements classified as unsafe, as well as measurements rejected by the current-based quality filter, were removed before training the offline forecasting model. Figure~\ref{fig:bad_data} summarizes the discarded data over the course of the campaign. Most rejected points occur during Phase~2 and are associated with heavily degraded IV sweeps identified by the current-based filter, whereas only a single unsafe measurement was observed during Phase~1, as previously discussed. The resulting curated dataset was then used for offline modeling with the KWW-based GP introduced in this section, enabling both interpolation within the measured domain and short- and long-horizon forecasting of the device response at a target voltage. Although the examples below focus on \SI{-1.5}{\volt}, the same framework can be applied to any target voltage within the range of the measured IV sweeps.

\begin{figure}[!t]
    \centering
    \includegraphics[width=\columnwidth]{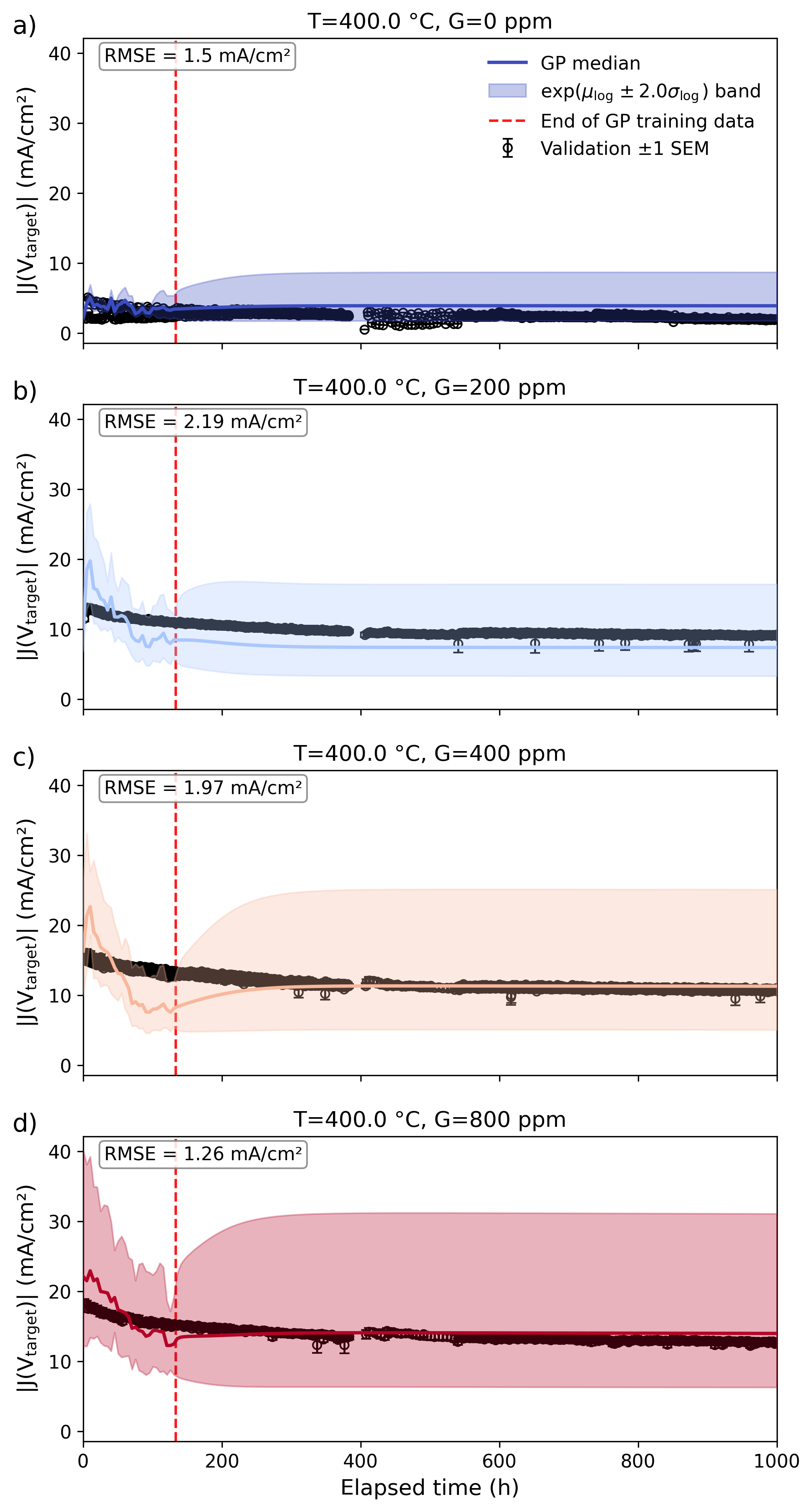}
    \caption{Time-dependent validation of the offline KWW-based GP model for the device response at \(V_{\mathrm{target}}=-1.5~\mathrm{V}\), shown at \(T=400~^\circ\mathrm{C}\) for four hydrogen concentrations: a) 0, b) 200, c) 400, and d) 800 ppm.}
    \label{fig:gp_validation_03_19}
\end{figure}

Figure~\ref{fig:gp_validation_03_19} compares the long-horizon predictions of the KWW-based GP with the auxiliary validation dataset (Fig.~\ref{fig:validation_dataset}). The GP was trained only on the initial \(\sim 133\) h of SAL-generated data, as indicated by the vertical dashed red line, and the remaining time window therefore probes extrapolation rather than interpolation. Figure~\ref{fig:gp_validation_03_19} shows that, despite being trained only on the first \(\sim 133\) h, the KWW-based GP captures the main long-time structure of the validation trajectories across all four hydrogen concentrations. In all cases, the model reproduces the qualitative pattern of a relatively rapid early transient followed by gradual flattening toward a concentration-dependent long-time level, which is precisely the type of behavior that motivated the use of a saturating KWW prior mean. The ordering of the trajectories with increasing H\textsubscript{2} concentration is also largely preserved, with larger concentrations corresponding to larger \(|I(V_{\mathrm{target}})|\) throughout most of the time window.

The extrapolation quality is not uniform across all conditions, which is expected for a forecasting problem of this type. At \(G=0~\mathrm{ppm}\), the response remains low and only weakly time dependent, and the GP median tracks that nearly stationary behavior well. At \(G=400\) and \(800~\mathrm{ppm}\), the model also reproduces the observed long-time plateau reasonably well, indicating that the structured mean captures the dominant degradation trend even well beyond the training horizon. The largest mismatch occurs at \(G=200~\mathrm{ppm}\), where the model tends to underpredict the late-time current, consistent with the larger RMSE reported in the panel. Even in that case, however, the forecast remains in the correct current range and preserves the overall saturating time dependence.

Another important feature of Fig.~\ref{fig:gp_validation_03_19} is the behavior of the uncertainty bands beyond the red dashed line. As expected, the predictive interval broadens once the model leaves the training region, reflecting increasing epistemic uncertainty in long-horizon extrapolation. Importantly, the uncertainty grows around a physically plausible saturating trajectory rather than around a linearly drifting or mean-reverting forecast. This behavior supports the use of the KWW-based mean as a practical prior for long-horizon prediction: it does not eliminate extrapolation uncertainty, but it regularizes that uncertainty toward degradation curves that remain physically interpretable over long times.

An important limitation of long-horizon forecasting based on SAL data arises in regions that are sparsely sampled over time. For a given \((T,G)\) region, insufficient temporal coverage limits the ability of the KWW-based model to accurately capture the degradation dynamics and, consequently, reduces forecasting reliability. Predictions should therefore be interpreted in the context of the sampling density and temporal coverage available for the specific \((T,G)\) region being forecasted.

Although the KWW model performed well in the present study, it should not be interpreted as the unique GP formulation for this problem. Alternative mean--covariance decompositions could also be successful,
provided that they preserve physically plausible long-time behavior and allow condition-dependent residual structure. Examples include multi-term stretched-exponential means for two-stage degradation, changepoint or nonstationary kernels for regime transitions, and monotonicity-aware GP formulations when strictly decreasing trajectories are required. We therefore view our proposed kernel as a pragmatic and effective extrapolation model rather than a definitive degradation law: its main advantage is that it
combines a saturating long-time prior with a residual kernel flexible enough to capture multi-scale deviations across time, temperature, and gas concentration.

\section{Conclusions}
We introduced a Safe Active Learning (SAL) framework for autonomous reliability characterization of Ga\(_2\)O\(_3\)-based diode sensors under coupled thermal and hydrogen stress. SAL combines a GP model of time-varying rectification with a conservative time-window safety criterion, a trust region anchored to measured-safe conditions. It uses a two-phase sampling strategy that first explores the high-rectification region and then progressively relaxes the rectification target as the device degrades. Simulation results showed that this strategy can safely expand the explored domain while learning the evolving rectification surface. Experimental implementation on an automated high-temperature probe-station platform demonstrated that SAL can generate a curated, time-resolved IV dataset under real operating conditions while maintaining conservative behavior in Phase~1 and intentionally moving toward riskier conditions only in Phase~2.

We further showed that the SAL-generated dataset can serve as the basis for offline long-horizon response modeling. Because the local RBF\(+\)linear kernel used online by SAL is not well suited to long-time extrapolation, we introduced a structured GP with a condition-dependent KWW degradation mean and a residual covariance kernel. On an auxiliary long-duration validation dataset, this model captured the main saturating degradation trends and preserved the systematic ordering with hydrogen concentration over most of the time window, while appropriately reflecting increased uncertainty beyond the training horizon. Taken together, these results suggest that safety-aware autonomous characterization can reduce manual experimentation burden while producing datasets that are directly useful for subsequent degradation forecasting.

Although the present study focuses on a rectifying Ga\(_2\)O\(_3\)-based device and uses rectification as the safety observable, the SAL framework is not restricted to this specific system. More generally, it should be applicable to other devices and experimental platforms whenever a physically meaningful safety metric can be defined and measured in situ during autonomous operation. 

\section*{Data availability}
Data will be made available upon request.

\section*{Author contributions}
D.F. drafted the manuscript with input from all co-authors, designed the algorithm, implemented it, and performed data analysis. A.Z. contributed to data analysis and assisted in the conceptualization of the workflow, including analytical procedures. W.C. supported data analysis, experimental setup, and equipment troubleshooting. A.S. fabricated all devices used in this work. The authors acknowledge the use of OpenAI’s ChatGPT and GitHub Copilot to assist with the implementation of code sections based on author-defined software architectures and designs, and with grammar and typographical review of the manuscript. All generated code was reviewed and verified by the authors prior to execution, and all conceptual content, analyses, and interpretations provided by the model were independently validated by the authors. The authors retain full responsibility for the scientific conclusions presented in this work. All authors approved the final version of the manuscript.

\section*{Conflicts of interest}
There are no conflicts to declare.

\section*{Acknowledgements}
This work was authored by the National Laboratory of the Rockies for the U.S. Department of Energy (DOE), operated under Contract No. DE-AC36-08GO28308. Funding for the SAL algorithm development and autonomous probe station measurements was provided by the Office of Critical Minerals and Energy Innovation (CMEI), Advanced Materials \& Manufacturing Technologies program (AMMTO). Funding for AI-assisted model development for a long-term reliability forecasting was provided by Office of Science (SC), Basic Energy Sciences (BES). The views expressed in the article do not necessarily represent the views of the DOE or the U.S. Government. The U.S. Government retains and the publisher, by accepting the article for publication, acknowledges that the U.S. Government retains a nonexclusive, paid-up, irrevocable, worldwide license to publish or reproduce the published form of this work, or allow others to do so, for U.S. Government purposes.





\appendix

\setcounter{section}{0}
\setcounter{figure}{0}
\setcounter{table}{0}
\setcounter{equation}{0}

\renewcommand{\thesection}{A\arabic{section}}
\renewcommand{\thefigure}{A\arabic{figure}}
\renewcommand{\thetable}{A\arabic{table}}
\renewcommand{\theequation}{A\arabic{equation}}

\section{Mathematical formulation and implementation of the SAL algorithm}
\label{A1}
\subsection{Rectification as a safety observable}

At each chosen \((T,G)\), the platform records one or more IV sweeps. Rather than evaluating rectification from a single interpolated point at \(\pm V_0\), SAL computes it from an intra-band comparison around the target voltage magnitude \(\lvert V_0\rvert\). This procedure is
illustrated in Fig.~\ref{fig:iv_sample}. For each sweep, only data within the voltage band 
\begin{equation}
\mathcal{B}(V_0)
=
\left\{
V:\;
\lvert \lvert V\rvert - \lvert V_0\rvert \rvert \le 0.1~\mathrm{V}
\right\}
\label{eq:rect_band}
\end{equation}
are retained. The filtered data are then split into positive- and
negative-bias subsets,
\begin{equation}
\mathcal{B}_{+}=\{V \in \mathcal{B}: V \ge 0\},
\qquad
\mathcal{B}_{-}=\{V \in \mathcal{B}: V < 0\}.
\end{equation}

SAL then performs an intra-band matching step: for each positive-bias point \(V_p \in \mathcal{B}_{+}\), it searches the negative-bias subset for the point \(V_n \in \mathcal{B}_{-}\) whose magnitude \(\lvert V_n\rvert\) is closest to \(\lvert V_p\rvert\). A pair is accepted only if the mismatch in voltage magnitude is smaller than a tolerance \(\delta_V\), chosen from the voltage spacing of the sweep.
For each accepted pair, SAL computes the pairwise rectification ratio
\begin{equation}
R_{\mathrm{pair}}
=
\frac{\lvert I(V_p)\rvert}{\max\!\bigl(\lvert I(V_n)\rvert,\epsilon\bigr)},
\label{eq:r_pair}
\end{equation}
where \(\epsilon\) is a small numerical floor introduced to avoid division by zero. Pairs with \(R_{\mathrm{pair}} < 1\) are discarded as non-physical for the intended rectifying regime.

This procedure yields a set of intra-band rectification ratios for each sweep. SAL then defines the sweep-level rectification as the median of the valid intra-band ratios,
\begin{equation}
R_{\mathrm{sweep}}
=
\mathrm{median}\!\left\{
R_{\mathrm{pair}}
\right\},
\label{eq:r_sweep}
\end{equation}
which provides robustness against local current noise and imperfect voltage matching inside the \(\pm V_0\) bands. If multiple sweeps are recorded at the same \((T,G)\), the condition-level rectification is
taken as the median across valid sweeps, 
\begin{equation}
R
=
\mathrm{median}\!\left\{
R_{\mathrm{sweep},1},
R_{\mathrm{sweep},2},
\dots
\right\},
\label{eq:r_condition}
\end{equation}
and the empirical spread is estimated from the standard deviation across valid sweep-level medians.

In parallel, SAL tracks how many sweeps produce valid intra-band rectification values. If too large a fraction of sweeps at a given \((T,G)\) fails the matching or validity checks, that condition is treated as unreliable and may be excluded from future sampling. The empirical spread is used for quality control and diagnostics only; it is not passed to the GP as a heteroscedastic noise term.

For the \(\mathrm{Ga_2O_3}\) diode, requiring
\begin{equation}
R \ge h
\end{equation}
at a specified \(\lvert V_0\rvert\) ensures that reverse leakage remains small compared to forward conduction at that bias. This provides a conservative, device-physics-motivated proxy for avoiding operating regimes that cause severe junction degradation.

\subsection{GP model and additive kernel}

The data set consists of time-stamped triplets
\begin{equation}
\begin{aligned}
  x_i &= (t_i, T_i, G_i), \\
  y_i &= \log R_i,
\end{aligned}
\end{equation}
with strictly increasing completion times \(t_1 < \dots < t_N\). SAL models the log-rectification surface as
\begin{equation}
  f(t,T,G) = \log R(t,T,G),
\end{equation}
with a Gaussian-process prior
\begin{equation}
  f \sim \mathcal{GP}\bigl(m(\cdot), k(\cdot,\cdot)\bigr).
\end{equation}
Inputs \((t,T,G)\) are normalized to \([0,1]^3\) using a BoTorch \texttt{Normalize} transform, and the scalar output is standardized after applying the logarithm using a \texttt{Standardize} outcome transform.

The covariance kernel \(k\) is chosen as an additive combination of a squared-exponential (RBF) term with automatic relevance determination (ARD) and a linear term,
\begin{equation}
  k(x,x') = k_{\mathrm{RBF}}(x,x') + k_{\mathrm{lin}}(x,x'),
  \label{SAL_kernel}
\end{equation}
where \(x = (t,T,G)\) and \(x' = (t',T',G')\). The RBF component uses an ARD lengthscale for each input dimension, with a log-normal prior on the lengthscales (scaled with input dimensionality) and a hard lower bound of \(0.025\) for numerical stability. The linear component is defined over all input dimensions and captures approximately affine trends in time, temperature, and gas concentration that are not well represented by purely local kernels, and the mean function \(m\) is a constant mean shared across the domain. Overall, the covariance structure is the sum of BoTorch's \texttt{SingleTaskGP} kernel and an additional linear term, allowing the surrogate to model both smooth local variations and global first-order trends in \((t,T,G)\).

In the implementation, we do not supply per-point observation variances; instead, the GP learns a single Gaussian noise variance, subject to a minimum noise level. Sweep-level variability is therefore absorbed into a homoscedastic noise term. For any test input \(x\), the GP posterior over \(f(x)\) is Gaussian,
\begin{equation}
  f(x) \mid \mathcal{D} \sim \mathcal{N}\bigl(\mu_f(x),\, \sigma_f^2(x)\bigr),
\end{equation}
where \(\mathcal{D}\) denotes all data collected so far. The GP hyperparameters are re-estimated at each iteration by maximizing the marginal likelihood in BoTorch.

\subsection{Adaptive completion-time window}

Because the completion time of each experiment is not known at planning time, SAL maintains a history of observed durations (in minutes) between scheduling and completion. Let
\begin{equation}
  d_j = t^{(\mathrm{comp})}_j - t^{(\mathrm{plan})}_j
\end{equation}
denote the duration of experiment \(j\). From the last \(L_H\) durations, SAL constructs an empirical high quantile
\begin{equation}
  H_{\mathrm{raw}} = Q_{p_H}\bigl(\{d_j\}_{j = N-L_H+1}^{N}\bigr),
\end{equation}
where \(Q_{p_H}\) is the empirical quantile at level \(p_H\) (for example, \(p_H = 0.9\)). A safety factor \(c_H > 1\) and clipping bounds \(H_{\min}\), \(H_{\max}\) are then applied:
\begin{equation}
  H = \min\bigl\{ H_{\max},\, \max\{ H_{\min},\, c_H H_{\mathrm{raw}} \} \bigr\}.
\end{equation}
This clipping prevents extreme horizon expansion due to rare long-duration outliers while ensuring a conservative planning window. Given the current time \(t_{\mathrm{now}}\), the algorithm defines a discrete completion-time window
\begin{equation}
\begin{aligned}
  \mathcal{T}_{\mathrm{win}}(t_{\mathrm{now}}) 
  &= \bigl\{\, t_{\mathrm{now}} + \Delta t_m : \Delta_{\min} \le \Delta t_m \le H, \\
  &\quad m = 1,\dots,M \,\bigr\},
\end{aligned}
\end{equation}
which enumerates plausible completion times for an experiment launched at \(t_{\mathrm{now}}\). \(\Delta_{\min} > 0\) enforces strictly increasing completion times and prevents degenerate zero-offset evaluations. All subsequent safety decisions are made against this entire completion-time window rather than a single nominal completion time.

\subsection{Log-normal LCB and time-window-safe set}

A user-defined rectification threshold \(h\) (for example, \(h = 50\)) specifies the minimum acceptable rectification. Because the GP is defined in log space, the marginal posterior at \((t,T,G)\) is
\begin{equation}
  f(t,T,G) \mid \mathcal{D} \sim 
  \mathcal{N}\bigl(\mu_f(t,T,G),\, \sigma_f^2(t,T,G)\bigr).
\end{equation}
The corresponding rectification
\begin{equation}
  R(t,T,G) = \exp\!\bigl(f(t,T,G)\bigr)
\end{equation}
is therefore approximated as log-normal, \(R \sim \mathrm{LogNormal}(\mu_f,\sigma_f^2)\). The \(\alpha\)-quantile of this distribution is
\begin{equation}
  Q_{\alpha}\bigl(R(t,T,G)\bigr)
  = \exp\!\bigl(\mu_f(t,T,G) + \sigma_f(t,T,G)\,\Phi^{-1}(\alpha)\bigr),
\end{equation}
where \(\Phi^{-1}\) is the standard normal quantile function.

SAL parameterizes the lower confidence bound (LCB) in terms of a standard-deviation multiplier \(\beta > 0\) and defines
\begin{equation}
  \ell_\beta(t,T,G) = \mu_f(t,T,G) - \beta\,\sigma_f(t,T,G),
\end{equation}
\begin{equation}
  L_\beta(t,T,G) = \exp\!\bigl(\ell_\beta(t,T,G)\bigr).
\end{equation}
By construction,
\begin{equation}
  \mathbb{P}\!\bigl(R(t,T,G) < L_\beta(t,T,G)\bigr)
  = \Phi(-\beta),
\end{equation}
so a larger \(\beta\) yields a more conservative bound. For example, with \(\beta = 3.5\),
\begin{equation}
  \Phi(-3.5) \approx 2.3\times 10^{-4},
\end{equation}
so the probability that the true rectification at a \emph{single} \((t,T,G)\) falls below its LCB is about \(0.023\%\) under the GP model.

To fold completion-time uncertainty into the safety decision, SAL aggregates these pointwise LCBs over the completion-time window. For each \((T,G)\), it computes
\begin{equation}
  L_{\mathrm{win}}(T,G)
  = \mathrm{Quantile}_{q}\bigl\{ L_\beta(t, T, G) : t \in \mathcal{T}_{\mathrm{win}}(t_{\mathrm{now}}) \bigr\},
\end{equation}
with \(q = 0.05\) corresponding to a requirement that at least \(95\%\) of the candidate completion times have LCBs above \(L_{\mathrm{win}}(T,G)\). The time-window-safe set at planning time \(t_{\mathrm{now}}\) is then
\begin{equation}
  \mathcal{S}_{\mathrm{safe}}(t_{\mathrm{now}}) =
  \bigl\{ (T,G) : L_{\mathrm{win}}(T,G) \ge h \bigr\}.
\end{equation}
If \(\mathcal{S}_{\mathrm{safe}}(t_{\mathrm{now}})\) is empty, SAL relaxes the confidence parameter \(\beta\) multiplicatively,
\begin{equation}
  \beta \rightarrow \max\{\beta_{\min},\, \gamma \beta\},
\end{equation}
with \(0 < \gamma < 1\), until either the safe set becomes non-empty or a prescribed maximum number of relaxation steps is reached. In this work, we set $\gamma=0.85$.

A simple probabilistic interpretation over the time window can be obtained by treating the values \(R(t_m,T,G)\) at the \(M\) candidate completion times \(t_m \in \mathcal{T}_{\mathrm{win}}\) as approximately independent. With \(\beta = 3.5\), the probability that any \emph{one} of them falls below its LCB is \(\Phi(-3.5) \approx 2.3\times 10^{-4}\). The probability that at least one among \(M\) such points violates its LCB is then approximated by
\begin{equation}
  p_{\mathrm{any}} \approx 1 - \bigl(1 - \Phi(-\beta)\bigr)^M.
\end{equation}
For \(M = 20\) and \(\beta = 3.5\),
\begin{equation}
  p_{\mathrm{any}} \approx 1 - (1 - 2.3\times 10^{-4})^{20}
  \approx 4.6\times 10^{-3},
\end{equation}
that is, an aggregate violation probability of order \(0.5\%\) over the entire completion-time window.

\subsection{Trust region in \texorpdfstring{\((T,G)\)}{(T,G)}}
Let
\begin{equation}
  \mathcal{S}_{\mathrm{meas}}(\tau)
  =
  \bigl\{ (T_i, G_i) : R_i \ge \tau \bigr\}
\end{equation}
denote the set of experimentally observed conditions that satisfy the current rectification threshold \(\tau\) (with \(\tau = h\) in Phase~1 and \(\tau = \tau_k\) in Phase~2).

Given iteration-dependent trust-region half-widths \(\Delta T_k\) and \(\Delta G_k\), SAL defines
\begin{equation}
\begin{aligned}
  \mathcal{T}_{\mathrm{TR},k}
  =
  \bigcup_{(T_s,G_s) \in \mathcal{S}_{\mathrm{meas}}(\tau)}
  \bigl\{ (T,G) :
    \lvert T - T_s\rvert \le \Delta T_k, \\
  \qquad\qquad\qquad\quad
    \lvert G - G_s\rvert \le \Delta G_k
  \bigr\}.
\end{aligned}
\end{equation}

During early post-seed iterations, the half-widths are kept intentionally tight to avoid aggressive extrapolation. After a prescribed warm-up period, they are relaxed to larger nominal values, enabling broader exploration while remaining anchored to empirically verified safe conditions.

The trust-region-gated safe set is then
\begin{equation}
  \mathcal{S}_{\mathrm{safe,TR}}(t_{\mathrm{now}})
  =
  \mathcal{S}_{\mathrm{safe}}(t_{\mathrm{now}})
  \cap
  \mathcal{T}_{\mathrm{TR},k}.
\end{equation}

\subsection{Phase~1: time-window-safe exploration at fixed rectification}

Phase~1 operates at a fixed rectification threshold \(h\) and aims to map the high-rectification region of the \(\mathrm{Ga_2O_3}\) device while honoring both the completion-time window and the trust-region constraints. For each \((T,G) \in \mathcal{S}_{\mathrm{safe,TR}}(t_{\mathrm{now}})\), SAL first aggregates predictive uncertainty over the completion-time window using a CVaR-style statistic. Let
\begin{equation}
  s_{\mathrm{CVaR}}(T,G)
  = \frac{1}{K_\alpha}
    \sum_{t \in \mathcal{T}_{\alpha}(T,G)} \sigma(t,T,G),
\end{equation}
where \(\mathcal{T}_{\alpha}(T,G)\) contains the top \(\alpha\)-fraction (e.g.\ 20\%) of predictive standard deviations over the window and \(K_\alpha\) is the number of such time samples. This emphasizes high-uncertainty completion times while remaining robust to outliers.

SAL then constructs a composite acquisition function
\begin{equation}
\begin{aligned}
  a(T,G)
  =\;&
  w_{\mathrm{unc}}\,U(T,G)
  + w_{\mathrm{explore}}\,D(T,G) \\
  &+ w_{\mathrm{revisit}}\,R_s(T,G)
  - w_{\mathrm{wall}}\,P_{\mathrm{wall}}(T,G).
\end{aligned}
\end{equation}
where \(U(T,G)\) is a normalized version of the CVaR-style time-aggregated uncertainty 
\(s_{\mathrm{CVaR}}(T,G)\). After computing \(s_{\mathrm{CVaR}}\) over the completion-time window, SAL performs a min–max normalization over the full \((T,G)\) grid,
\[
U(T,G)
=
\frac{s_{\mathrm{CVaR}}(T,G) - s_{\min}}
     {s_{\max} - s_{\min} + \varepsilon},
\]
where \(s_{\min}\) and \(s_{\max}\) are the minimum and maximum values of \(s_{\mathrm{CVaR}}\) across all candidate grid cells and \(\varepsilon\) is a small numerical constant. Thus \(U(T,G)\in[0,1]\), with the most time-uncertain condition assigned value 1 and the least uncertain assigned 0. This ensures that the uncertainty contribution is scale-invariant and comparable across iterations.

The term \(D(T,G)\) is a diversity weight encouraging sampling away from previously evaluated \((T,G)\) conditions. For each candidate point, SAL computes the Euclidean distance to the nearest previously sampled condition,
\[
d_{\min}(T,G) = \min_i \| (T,G) - (T_i,G_i) \|_2,
\]
and rescales this distance by
\[
r_{\mathrm{scale}}
=
\frac{1}{2}
\left(
|\Delta T_{\mathrm{TR}}|
+
|\Delta G_{\mathrm{TR}}|
\right),
\]
where \(\Delta T_{\mathrm{TR}}\) and \(\Delta G_{\mathrm{TR}}\) are the current trust-region half-widths.
The normalized distance
\(\tilde d(T,G)=\mathrm{clip}(d_{\min}/r_{\mathrm{scale}},0,1)\)
is then mapped to \([0,1]\) using a cubic smoothstep function. Consequently, \(D(T,G)\) is close to zero near existing samples and approaches one in spatially under-explored regions. Cells explicitly scheduled for time-based follow-up revisits are exempted from this repulsion so that temporal monitoring is not suppressed by proximity-based penalties.

The revisit score \(R_s(T,G)\) promotes periodic re-measurement of previously visited grid cells in order to track time-dependent drift. Let \(t_{\mathrm{last}}(T,G)\) denote the most recent visit time of a grid cell. SAL computes the elapsed time since the last visit,
\[
\Delta t_{\mathrm{age}}(T,G)
=
t_{\mathrm{now}} - t_{\mathrm{last}}(T,G),
\]
and converts it into a bounded priority score,
\[
R_s(T,G)
=
\begin{cases}
\mathrm{clip}\!\left(\dfrac{\Delta t_{\mathrm{age}}(T,G)}{t_{1/2}},\,0,1\right),
& \text{if the cell has been visited},\\[6pt]
0, & \text{otherwise}.
\end{cases}
\]
Here \(t_{1/2}\) is a user-defined half-life parameter controlling how quickly revisit priority saturates: immediately after a visit \(R_s\approx 0\), while if a condition has not been revisited for at least \(t_{1/2}\) minutes, the score saturates at 1. The revisit score is forcibly set to zero once a maximum number of revisits per cell has been reached. In addition, mild region-dependent scaling may reduce revisit pressure in benign low-\(T\), low-\(G\) regions and increase it in harsher operating regimes. 

Finally, \(P_{\mathrm{wall}}(T,G)\in[0,1]\) penalizes proximity to global domain boundaries. The penalty increases smoothly as \((T,G)\) approaches temperature limits or the upper gas-concentration boundary and decays toward zero in the interior of the domain, discouraging persistent edge-hugging behavior while still allowing boundary sampling when justified by uncertainty or safety considerations.

Outside the trust-region-gated safe set, and at permanently banned conditions, the acquisition is set to \(-\infty\).
The next condition is selected as
\begin{equation}
  (T_{\mathrm{next}}, G_{\mathrm{next}}) =
  \arg\max_{(T,G)} a(T,G),
\end{equation}
the experiment at \((T_{\mathrm{next}}, G_{\mathrm{next}})\) is executed, the rectification \(R_{\mathrm{meas}}\) is extracted from the IV data, and the actual completion time \(t_{\mathrm{actual}}\) is recorded. If the rectification measurement is invalid, or if the fraction of usable sweeps is below a specified threshold, \((T_{\mathrm{next}}, G_{\mathrm{next}})\) is added to the ban list and a counter of consecutive invalid experiments is incremented; the campaign is terminated if this counter exceeds a user-defined limit. Otherwise, the duration statistics are updated and SAL proceeds to the next iteration.

For valid measurements, SAL records whether \(R_{\mathrm{meas}} < h\). These events are counted as measured safety violations for post hoc analysis, while the online safety gate is enforced through the conservative lower bounds \(L_{\mathrm{win}}\).

\subsection{Phase~1 rescue procedure}

If \(\mathcal{S}_{\mathrm{safe}}(t_{\mathrm{now}})\) remains empty despite relaxation of \(\beta\), the GP may be pessimistic when extrapolating over the time domain, and thus SAL invokes a rescue routine. The algorithm selects the most recent \((T^\star, G^\star)\) with \(R \ge h\) and performs several additional experiments at this fixed condition, appending the resulting measurements to the data set.

The rescue data are then used to classify the situation:
\begin{itemize}
  \item If the majority of rescue measurements satisfy \(R \ge h\), the disappearance of \(\mathcal{S}_{\mathrm{safe}}\) is interpreted as a modelling artefact; Phase~1 is resumed with the augmented data.
  \item If the rescue measurements are mixed (both above and below \(h\)), \((T^\star,G^\star)\) is treated as lying near the safety boundary; Phase~1 is terminated and Phase~2 is started.
  \item If most rescue measurements are unsafe or become invalid, the system is considered to have drifted into an unreliable regime, and the campaign is terminated.
\end{itemize}
If rescue experiments at \((T^\star, G^\star)\) themselves yield invalid data, this condition is banned and the run is stopped rather than attempting to rebuild a safe set around an unstable operating point. These decisions are designed to be conservative, i.e., preserve the device integrity.

\subsection{Phase~2: exponential relaxation of the rectification target}

Phase~2 starts from the data accumulated during Phase~1 and gradually relaxes the rectification target from an initial value \(\tau_{\mathrm{start}}\) down to a final value \(\tau_{\mathrm{final}} \approx 1\), which means the device has degraded and is not rectifying anymore (resistor-like behavior). At iteration \(k\) of Phase~2, the current target is
\begin{equation}
  \tau_k = \tau_{\mathrm{final}} +
           \bigl(\tau_{\mathrm{start}} - \tau_{\mathrm{final}}\bigr)\,
           2^{-k / T_{1/2}},
           \label{tau_schedule}
\end{equation}
where \(\tau_{1/2}\) is a user-chosen half-life parameter.

Each Phase~2 iteration reuses the machinery of Phase~1, with the fixed threshold \(h\) replaced by \(\tau_k\). Three regimes are distinguished:
\begin{itemize}
  \item \emph{Safe mode} (\(\tau_k > 1 + \epsilon\), $\epsilon=0.001$): recompute the time-window-safe set at level \(\tau_k\), intersect with the trust region, and select the next \((T,G)\) by maximizing \(a(T,G)\) as in Phase~1. If the safe set remains empty after \(\beta\)-relaxation, switch to a trust-region-gated uncertainty fallback.
  \item \emph{Trust-region-gated uncertainty fallback}: when no \((T,G)\) is time-window-safe at level \(\tau_k\), ignore the lower confidence bound in the selection rule, restrict attention to the union of trust-region rectangles induced by previously measured-safe points (with respect to \(\tau_k\)), and choose the \((T,G)\) that maximizes the CVaR-style uncertainty statistic \(s_{\mathrm{CVaR}}(T,G)\). This encourages sampling near the empirical safety boundary without exploring arbitrary parts of the domain.
  \item \emph{Pure uncertainty mode} (\(\tau_k \approx 1\)): once \(\tau_k\) approaches the minimum rectification of interest, drop rectification-based safety gating altogether and maximize \(s_{\max}(T,G)\) over the entire \((T,G)\) grid (excluding banned points), refining the model in low-rectification regions.
\end{itemize}

As in Phase~1, each selected \((T,G)\) is evaluated, the data set and duration statistics are updated, and the algorithm records whether \(R_{\mathrm{meas}}\) lies below either the original Phase~1 threshold \(h\) or the current target \(\tau_k\). In this way, SAL provides a time-aware, rectification-based safe exploration of the operating space of the device under test. Fig.~\ref{fig:pseudocode} shows the high-level pseudocode of the SAL algorithm, summarizing its main steps and two-phase structure.

Because SAL enforces safety constraints at every queried (T,G) condition, the resulting campaign yields a curated set of IV curves collected over time within a certified safe operating region. In practice, this means that a successful SAL run on the real instrumentation “for free” produces a time-resolved IV dataset from which we can subsequently extract diode parameters, model device current, and analyze degradation mechanisms offline, without requiring additional dedicated experiments—substantially accelerating device characterization and lifetime studies.

\section{Simulations}
\label{A2}

\subsection{Simulated rectification surface}
\label{ground_truth}
In simulation mode, the rectification factor is generated from the general form
\begin{align}
  R_{\mathrm{sim}}(t,T,G)
  &= 1 + 99 \exp\!\Bigl[
      -\,k \Bigl(
          \alpha\,\max(T-T_0,0)
          \notag \\[-1mm]
  &\hphantom{= 1 + 99 \exp\Bigl[ -\,k \Bigl(}
          {}+ \beta\,\max(G,0)
          + \gamma\,\max(t,0)
      \Bigr)
    \Bigr]
    \label{eq:ground_truth}
\end{align}

The constants used in this work are  
$$
T_0 = 300~^\circ\mathrm{C}, \quad
\alpha = \frac{0.10}{300}, \quad
\beta = \frac{0.10}{1000}, \quad
\gamma = 10^{-4}, \quad
k = 41.82,
$$

This form ensures that the benign corner \((T\!\le\!300~^\circ\mathrm{C},G=0,t=0)\) yields \(R_{\mathrm{sim}}=100\), while progressively harsher conditions (increasing temperature, hydrogen concentration, or exposure time) produce monotonic decay toward values near unity. The exponential structure guarantees smoothness across the entire domain, while the use of non-negative offsets enforces that degradation only activates above a reference temperature of \(300~^\circ\mathrm{C}\) and for \(t>0\).

For fixed \((T,G)\), the model decays monotonically with time. Early in the campaign, \(R_{\mathrm{sim}}\) remains near its initial high value, followed by gradual attenuation as the effective exposure time increases. This reproduces the typical behavior of rectifying devices under extended thermal and gas stress. For fixed \(t\) and \(G\), rectification remains constant for \(T \le 300~^\circ\mathrm{C}\). Above this threshold, the term \(\max(T-300,0)\) introduces a temperature-activated reduction in rectification, with more rapid loss at higher temperatures. For fixed \(t\) and \(T\), increasing hydrogen concentration \(G\) produces monotonic reduction in \(R_{\mathrm{sim}}\). The model assigns zero penalty at \(G=0\), and the degradation rate increases smoothly with concentration, mimicking accelerated stress under reducing environments.

Each simulated observation can be corrupted by Gaussian measurement noise,
\[
  R_{\mathrm{obs}} \sim \mathcal{N}\!\bigl(
    R_{\mathrm{sim}},\,\sigma_{\mathrm{meas}}^{2}
  \bigr),
\]
with standard deviation
\[
  \sigma_{\mathrm{meas}}
  = \max\bigl(0.5,\;0.05\,R_{\mathrm{sim}}\bigr),
\]
and values are truncated below to enforce \(R_{\mathrm{obs}} \ge 1\).

\subsection{Simulation setup}

The SAL algorithm was executed in simulation mode over the search domain \(T \in (300,600)\,\si{\celsius}\) and \(G \in (0,1000)\,\text{ppm}\). The algorithm was initialized with four seed points: (\SI{310}{\celsius}, \SI{100}{ppm}),
(\SI{315}{\celsius}, \SI{200}{ppm}), (\SI{330}{\celsius}, \SI{50}{ppm}), and (\SI{340}{\celsius}, \SI{0}{ppm}). The campaign time \(t\) was initialized at zero, and each subsequent measurement was assigned a simulated completion time
\(t_{k+1}\) in minutes according to
\[
  t_{k+1} = t_k + \Delta t_{\min} + \varepsilon_k,
\]
with a base increment \(\Delta t_{\min} = \SI{5}{\minute}\) and Gaussian jitter \(\varepsilon_k \sim \mathcal{N}(0,\,\SI{1}{\minute}^2)\), truncated so that \(t_{k+1} - t_k \ge \SI{2.5}{\minute}\). The resulting durations
\(d_k = t_{k+1} - t_k\) were stored online, and the effective completion-time horizon was adapted as
\[
  t_{h,\mathrm{eff}} = 1.5 \times P_{90}\{d_k\},
\]
clipped to the interval \([\SI{0.1}{\minute}, \SI{60}{\minute}]\). In the absence of sufficient duration statistics at the beginning of the campaign, the nominal planning horizon was initialized to the user-specified value of \(\SI{20}{\minute}\). For each planning step at time \(t_{\mathrm{now}}\), SAL evaluated the simulated rectification on a discrete completion-time grid
\[
  t \in [t_{\mathrm{now}} + \Delta t_{\min},\, t_{\mathrm{now}} + t_{h,\mathrm{eff}}],
\]
with \(\Delta t_{\min} = \SI{5}{\minute}\) and time step
\(\Delta t = \SI{1}{\minute}\). The completion-time window was discretized with a \SI{1}{\minute} step, and this grid was then used to compute time-window lower confidence bounds and the associated time-window-safe set in the simulation.

In the \((T,G)\) space, SAL operated on a regular lattice with a temperature increment of \(\Delta T = \SI{5}{\celsius}\) and a gas-concentration increment of \(\Delta G = \SI{50}{ppm}\). The safety gate was defined using a log-LCB with confidence parameter \(\beta = 3.5\) and a rectification safety threshold \(h = 10\). The run comprised 100 iterations in Phase~1 (fixed threshold \(h\)) followed by 100 iterations in Phase~2, during which the rectification target was exponentially relaxed toward \(\tau_{\mathrm{final}} = 1\) with half-life parameter \(\tau_{1/2} = 3.8\), so that transition to pure uncertainty in phase 2 happens after 50 iterations of progressive relaxation of \(\tau\) in phase 2. A trust region in \((T,G)\) was enabled throughout: during an initial warm-up of three iterations, the trust-region half-widths were set to \(\Delta T_{\mathrm{TR}} = \SI{25}{\celsius}\) and \(\Delta G_{\mathrm{TR}} = \SI{100}{ppm}\), and then expanded to \(\Delta T_{\mathrm{TR}} = \SI{50}{\celsius}\) and \(\Delta G_{\mathrm{TR}} = \SI{200}{ppm}\) for the remainder of the run. During early post-seed iterations, the trust-region half-widths are kept intentionally tight to avoid aggressive extrapolation; after a prescribed warm-up period, the half-widths are relaxed to their nominal values. This staged enlargement of the trust region balances early conservatism with later exploratory flexibility.

\section{Mathematical formulation of the KWW kernel for long-horizon forecasting}
\label{A3}

Low-order functions of normalized temperature \(\tilde{T}\) and normalized hydrogen concentration were used to parameterize the terms of the KWW kernel:
\(\tilde{G}\):
\begin{align}
B(T,G)
&=
c + a_T \tilde{T} + a_{TT}\tilde{T}^2
  + a_G \tilde{G} + a_{GG}\tilde{G}^2,
\label{eq:kww_B}
\\[3pt]
A(T,G)
&=
\mathrm{softplus}\!\Big(
d_0 + d_T \tilde{T} + d_G \tilde{G}
+ d_{GG}\tilde{G}^2 + d_{TG}\tilde{T}\tilde{G}
\Big),
\label{eq:kww_A}
\\[3pt]
\lambda(T,G)
&=
\exp\!\Big(
r_0 + r_T \tilde{T} + r_G \tilde{G}
+ r_{TG}\tilde{T}\tilde{G}
\Big).
\label{eq:kww_lambda}
\end{align}
This form was chosen to encode the minimum structure needed for physically plausible extrapolation while remaining flexible enough to fit the data. The quadratic terms in Eq.~\eqref{eq:kww_B} allow the initial
response level to vary nonlinearly with operating condition. The \(\mathrm{softplus}\) transform in Eq.~\eqref{eq:kww_A} enforces \(A(T,G) > 0\), ensuring that the KWW term represents a drop from the
baseline rather than an unphysical sign change. The quadratic gas term in \(A(T,G)\) allows the degradation depth to vary nonlinearly with hydrogen concentration, which is useful when the response begins to saturate at higher gas exposure. Finally, the bilinear interaction terms \(\tilde{T}\tilde{G}\) in Eqs.~\eqref{eq:kww_A} and \eqref{eq:kww_lambda} allow temperature to amplify the effect of hydrogen concentration, reflecting the empirical expectation that harsher thermal conditions accelerate gas-assisted degradation rather than acting independently of gas concentration.

The stretch exponent \(\beta\) controls the degree of dispersion in the degradation kinetics. In the KWW model, \(\beta=1\) reduces to a simple exponential, whereas \(\beta<1\) produces the broader, long-tailed decay expected when multiple processes contribute to the measured response. To keep \(\beta\) in a physically reasonable range, we parameterize it as
\begin{equation}
\beta = 0.05 + 0.65\,\sigma(\eta_\beta),
\qquad
\beta \in (0.05,\,0.70),
\label{eq:kww_beta}
\end{equation}
where \(\sigma(\cdot)\) is the logistic sigmoid and \(\eta_\beta\) is an unconstrained latent parameter optimized during training. Thus, \(\eta_\beta\) itself has no direct physical interpretation; it is
introduced only to map the physical stretch exponent \(\beta\) into a bounded interval that favors sub-exponential, long-tailed degradation rather than near-exponential decay that would flatten too quickly at long times.

To further regularize long-horizon extrapolation, the offline model was trained with informative but non-restrictive hyperparameter priors and constraints in normalized input space. Specifically, weakly informative normal priors were placed on the coefficients defining the baseline, amplitude, and rate functions in Eqs.~\eqref{eq:kww_B}--\eqref{eq:kww_lambda}, while the latent parameter \(\eta_\beta\) was assigned a prior favoring smaller \(\beta\) values within the bounded interval of Eq.~\eqref{eq:kww_beta}. This biases the model toward sub-exponential, long-tailed degradation unless the data support faster saturation. In the residual GP, log-normal priors were placed on the kernel lengthscales, with a longer prior center along the time dimension for the global rational-quadratic term, and a small minimum inferred noise level was enforced for numerical stability. These choices do not fix the extrapolation behavior, but they regularize it toward physically plausible long-time degradation trajectories.

The KWW mean captures only the dominant long-time trend. Residual departures from this idealized degradation law are modeled by a GP with covariance
\begin{equation}
k_{\mathrm{res}}(x,x')
=
\sigma_k^2
\Big[
k_{\mathrm{RQ}}^{\mathrm{all}}(x,x')
+
k_{\mathrm{Mat}}^{t}(t,t')\,
k_{\mathrm{RQ}}^{TG}\!\big((T,G),(T',G')\big)
\Big].
\label{eq:kww_residual_kernel}
\end{equation}

The first term is a global rational-quadratic kernel over all input dimensions and captures multiscale residual structure that is not explained by the KWW mean. We prefer a rational-quadratic form here because it acts as a scale mixture of RBF kernels and is therefore better suited to residual behavior with more than one characteristic correlation length. The second term is a product kernel consisting of a Mat\'ern-\(5/2\) kernel in time and a rational-quadratic kernel over \((T,G)\). This term allows condition-specific temporal deviations from the global degradation trend, so that nearby operating conditions can share similar residual time evolution without forcing all conditions to follow exactly the same trajectory. The model is trained on \(\log |I(V_{\mathrm{target}})|\) as a function of elapsed time, temperature, and hydrogen concentration, using safe data provided by SAL. In this formulation, the KWW mean provides the long-horizon extrapolation prior, while the GP residual supplies local flexibility around that prior.

\bibliography{references} 
\bibliographystyle{rsc} 

\clearpage
\onecolumn
\newgeometry{margin=1in}
\setstretch{1.0}
\pagestyle{plain}

\begin{titlepage}
    \centering
    {\large\bfseries Electronic Supplementary Information (ESI)\par}
    \vspace{1.5cm}
    {\Large\bfseries Autonomous Reliability Qualification of Ga$_2$O$_3$-based Hydrogen and Temperature Sensors via Safe Active Learning\par}
    \vspace{2cm}
    {\large Davi Febba, William A. Callahan, Anna Sacchi, Andriy Zakutayev\par}
    \vfill
    {\large \today\par}
\end{titlepage}
\onecolumn

The Gaussian process (GP) models used in this work are trained with a logarithmic outcome transform applied to the response variable. Let $R$ denote the rectification factor and define the transformed variable

\begin{equation}
Z = \log R .
\end{equation}

The GP therefore models the posterior distribution of $Z$, which is Gaussian:

\begin{equation}
Z \mid x \sim \mathcal{N}(\mu_{\log}(x), \sigma_{\log}^2(x)),
\end{equation}

where $x=(t,T,G)$ denotes the input variables. Because the model is Gaussian in log space rather than in the original response variable $R$, the predictive distribution in the original units is not Gaussian. Instead, the induced distribution of $R = e^Z$ is approximately lognormal.

BoTorch returns posterior moments in the original response units after undoing the outcome transform. However, since the model is Gaussian in log space, uncertainty is naturally symmetric in $\log R$, not in $R$ itself. As a result, plotting a symmetric linear-space interval of the form $R_{\mathrm{mean}} \pm 2\sigma$ would impose a Gaussian interpretation in the original units that is inconsistent with the transformed model.

For visualization purposes, we therefore reconstruct the equivalent lognormal representation of the predictive distribution. From the linear-space mean $m$ and variance $s^2$ returned by BoTorch, the corresponding log-space parameters are obtained using the standard lognormal relationships

\begin{equation}
\sigma_{\log}^2 = \ln\!\left(1 + \frac{s^2}{m^2}\right),
\end{equation}

\begin{equation}
\mu_{\log} = \ln(m) - \frac{1}{2}\sigma_{\log}^2 .
\end{equation}

The central tendency and uncertainty interval are then visualized using the back-transformed log-space quantities

\begin{equation}
R_{\mathrm{median}} = e^{\mu_{\log}},
\end{equation}

\begin{equation}
R_{\mathrm{lower}} = e^{\mu_{\log}-2\sigma_{\log}}, \qquad
R_{\mathrm{upper}} = e^{\mu_{\log}+2\sigma_{\log}} .
\end{equation}

In this representation the central curve corresponds to the \textit{posterior median} of $R$, and the interval $\exp(\mu_{\log} \pm 2\sigma_{\log})$ represents the symmetric uncertainty band in log space mapped back to the original units. This interval is multiplicatively symmetric about the median and reflects the asymmetric uncertainty structure implied by the log-transformed GP.

For surface visualizations of uncertainty, we additionally report the multiplicative uncertainty factor

\begin{equation}
\exp(2\sigma_{\log}),
\end{equation}

which indicates the factor by which the response may vary above or below the median at approximately the two-standard-deviation level in log space.

This visualization approach ensures that the plotted uncertainty faithfully reflects the probabilistic structure of the GP model while preserving interpretability in the original response units.

\begin{figure*}[!t]
    \centering
    \includegraphics[width=\textwidth]{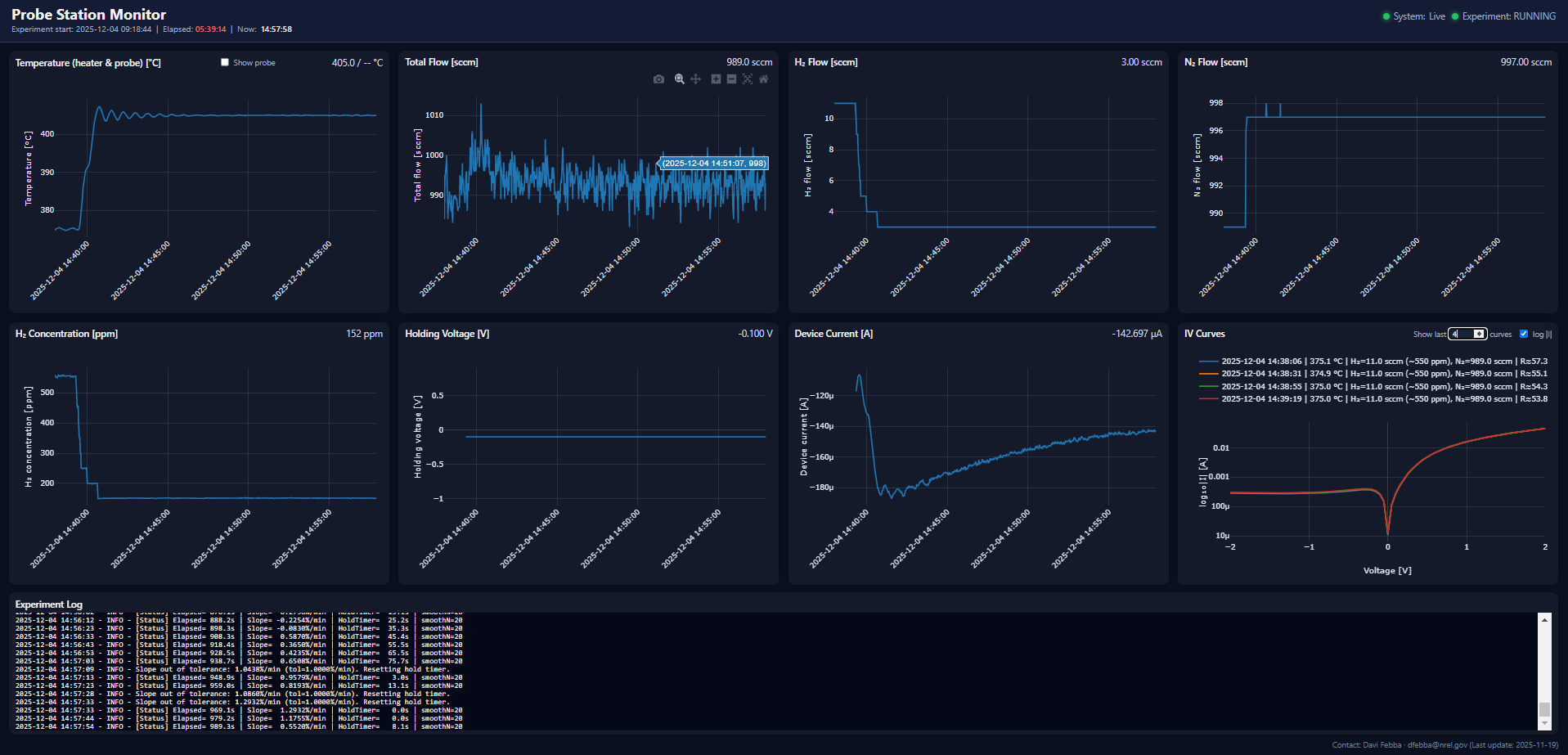}
    \caption*{\textbf{Fig. S1.} Overview of the monitoring interface developed in the study, showing data from a run of the SAL algorithm.}
\end{figure*}

\begin{figure*}[!t]
    \centering
    \includegraphics[width=\textwidth]{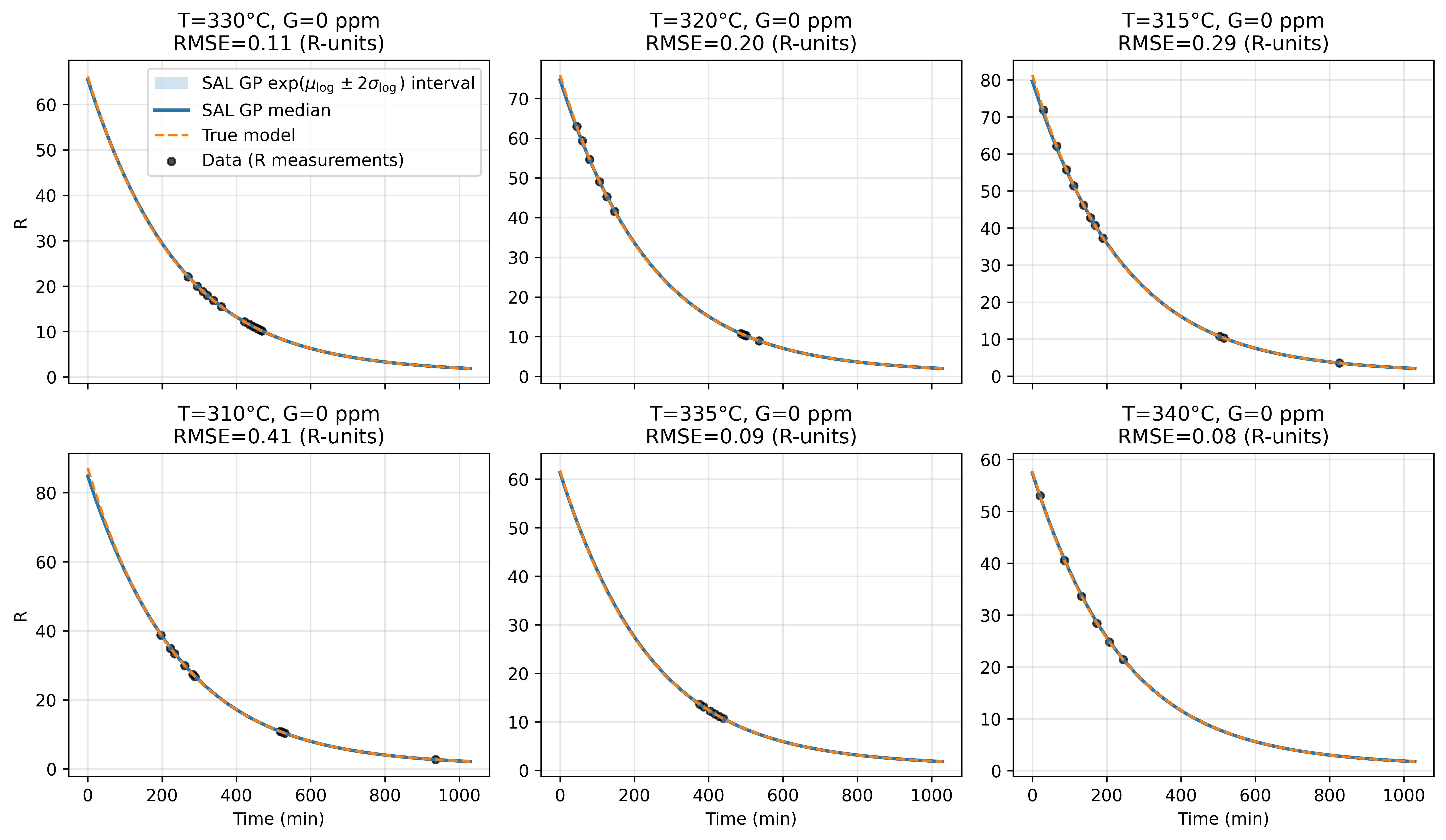}
    \caption*{Fig. S2. Comparison between the true model and the Gaussian process (GP) surrogate learned by SAL in regions of the search domain with strong data support. Each subplot shows the time evolution of the rectification factor $R$ at fixed $(T,G)$ conditions. The dashed orange line corresponds to the true model, the solid blue line to the SAL GP posterior median, and the shaded region indicates the GP uncertainty. Black markers show the $R$ measurements collected by SAL during the campaign. In these densely sampled regions, the GP predictions closely follow the true trajectories and the measured data, resulting in small RMSE values and narrow uncertainty bands. This agreement demonstrates that SAL accurately learns the underlying temporal behavior of $R$ when sufficient observations are available.}    
    \label{fig:simulations_partial_dependence_data}
\end{figure*}

\begin{figure*}[!t]
    \centering
    \includegraphics[width=\textwidth]{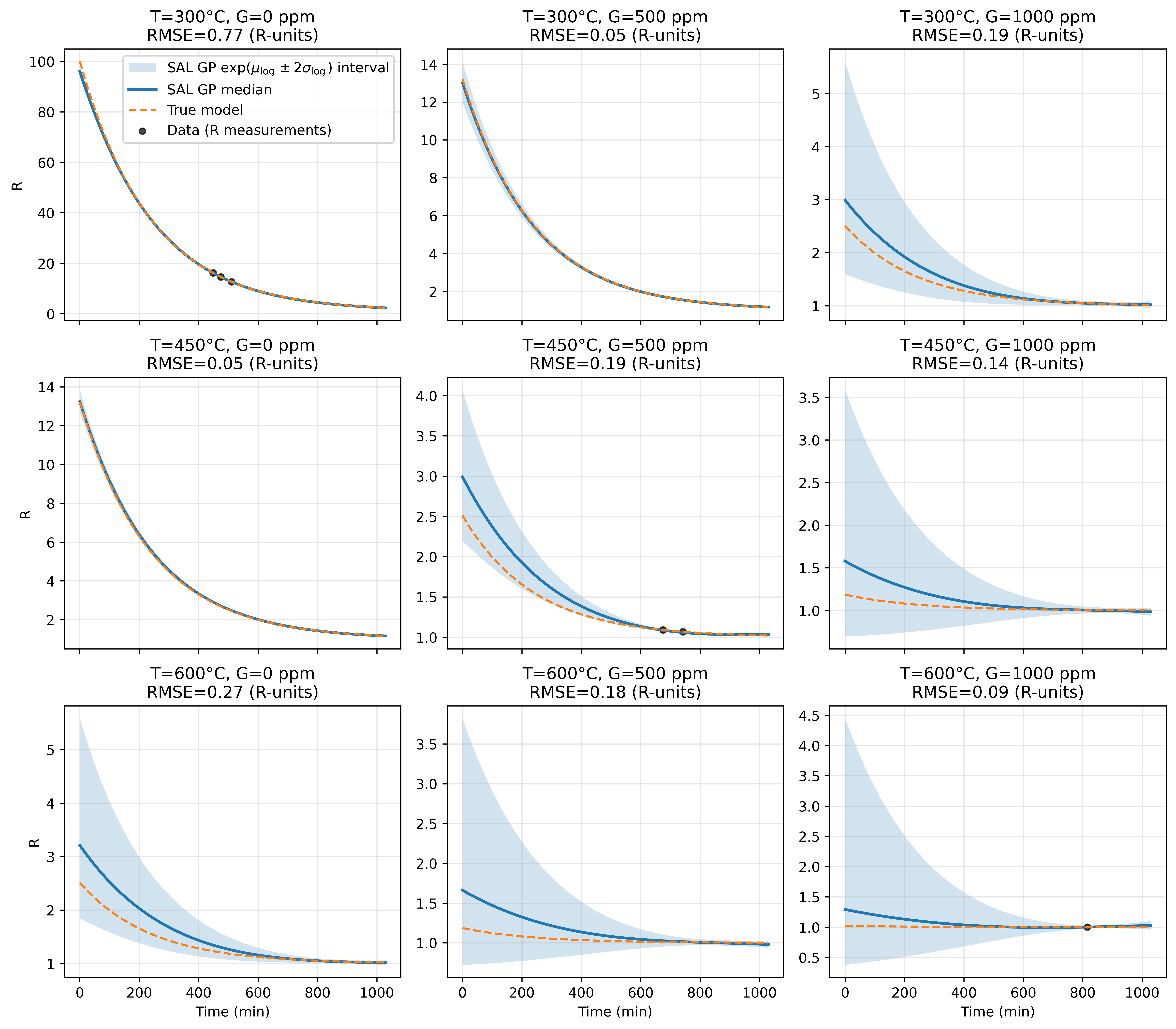}
    \caption*{Fig. S3. Comparison between the true model and the Gaussian process (GP) surrogate learned by SAL in regions of the search domain with limited or no direct data support. Each subplot shows the time evolution of the rectification factor $R$ at fixed $(T,G)$ conditions. The dashed orange line represents the true model, the solid blue line the SAL GP posterior median, and the shaded region the GP uncertainty. Even in sparsely sampled regions, the GP often reproduces the correct trends with small RMSE, indicating that the model successfully leverages correlations learned across time, temperature, and gas concentration to generalize beyond the sampled points. In cases where the condition lies further from the sampled region, the uncertainty increases and the RMSE can grow, reflecting the expected behavior of the GP during extrapolation.}   
    \label{fig:simulations_partial_depdendence}
\end{figure*}

\begin{figure*}[!t]
    \centering
    \includegraphics[width=\textwidth]{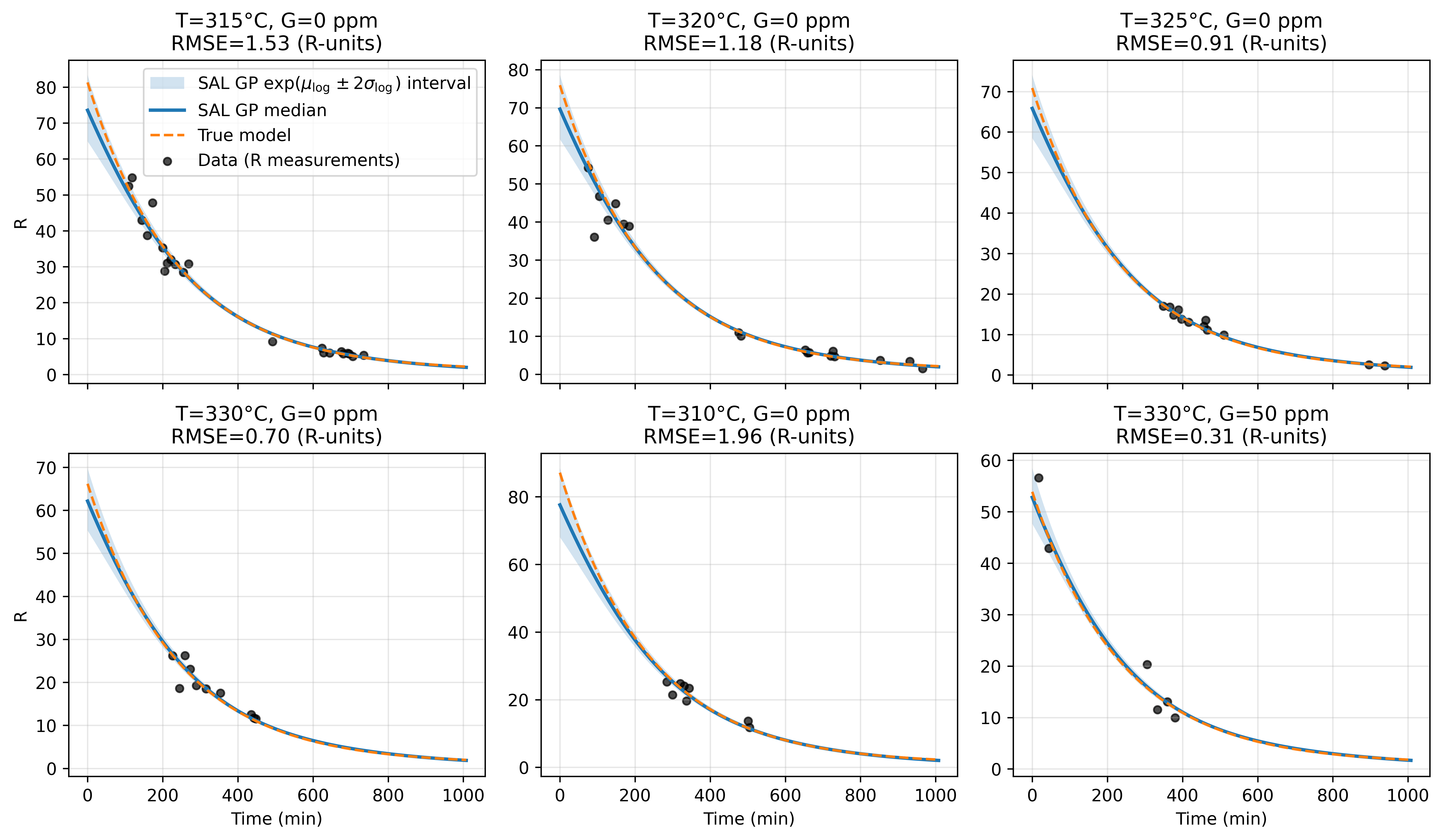}
    \caption*{Fig. S4. Comparison between the true model and the Gaussian process (GP) surrogate learned by SAL for time slices at several $(T,G)$ conditions in regions where SAL collected dense measurements. The dashed orange line corresponds to the true rectification model, the solid blue line to the GP posterior mean, and the shaded region indicates the GP uncertainty. Black markers represent the measured rectification factors obtained during the SAL campaign. Even with the higher measurement noise used in this simulation (10\% relative noise with a 0.5 absolute floor), the GP predictions closely follow the true trajectories in well-sampled regions, demonstrating that SAL can still recover the underlying model behavior despite increased experimental noise.
    }
\label{fig:simulations_partial_depdendence_data_noise}
\end{figure*}

\begin{figure*}[!t]
    \centering
    \includegraphics[width=\textwidth]{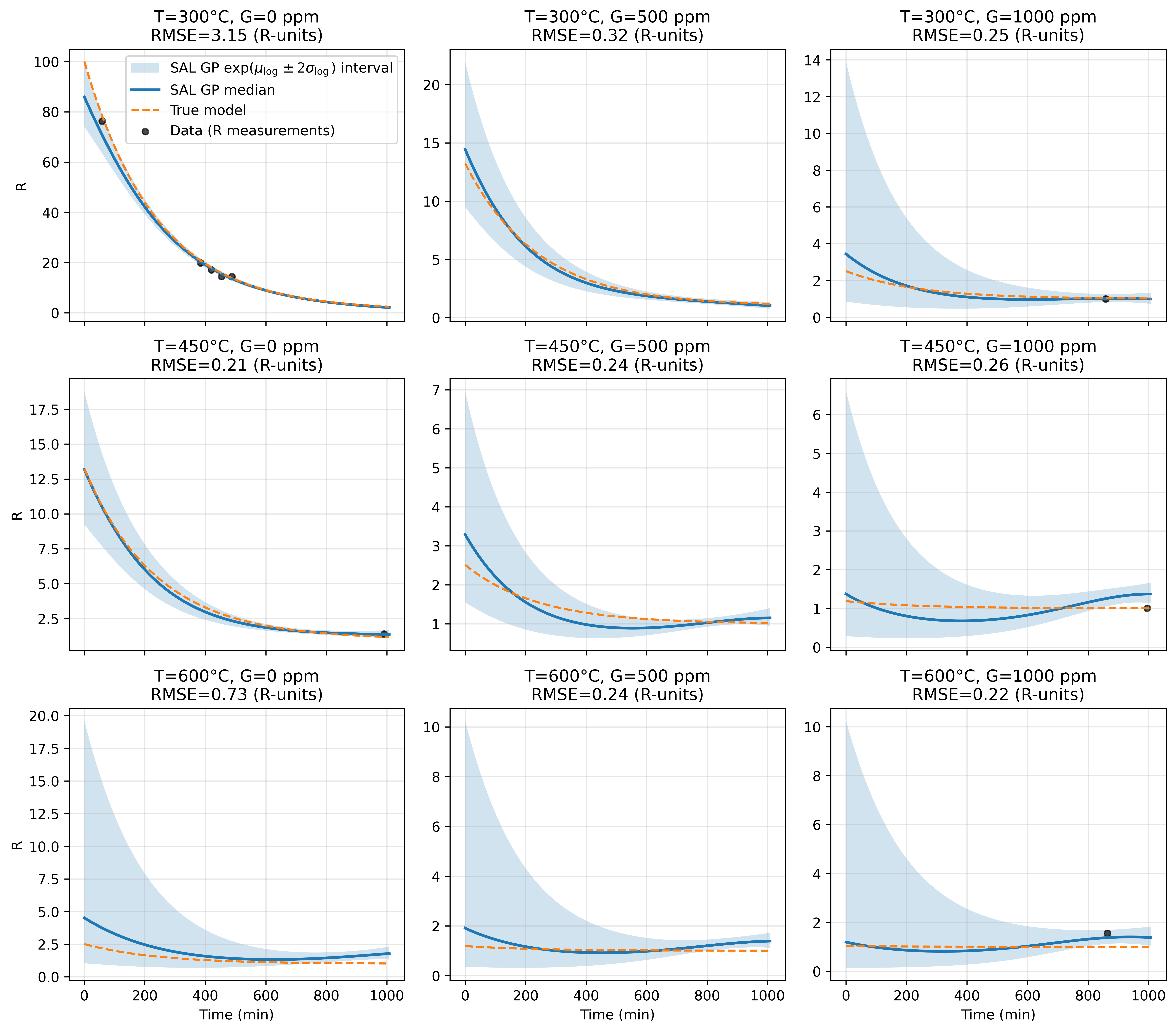}
    \caption*{Fig. S5. Comparison between the true model and the Gaussian process (GP) surrogate learned by SAL for time slices at several $(T,G)$ conditions in regions with limited direct data support. The dashed orange line represents the true model, the solid blue line the GP posterior mean, and the shaded region the GP uncertainty. Measurement points collected by SAL are shown as black markers when available. Because of the increased measurement noise used in this simulation (10\% relative noise with a 0.5 absolute floor), the GP uncertainty bands are wider and the RMSE is larger than in the noiseless case. Nevertheless, the GP predictions still capture the qualitative temporal trends and converge toward the true model where data become available.
    }
    \label{fig:simulations_partial_depdendence_data_noise}
\end{figure*}

\begin{figure*}[!t]
    \centering
    \includegraphics[width=\textwidth]{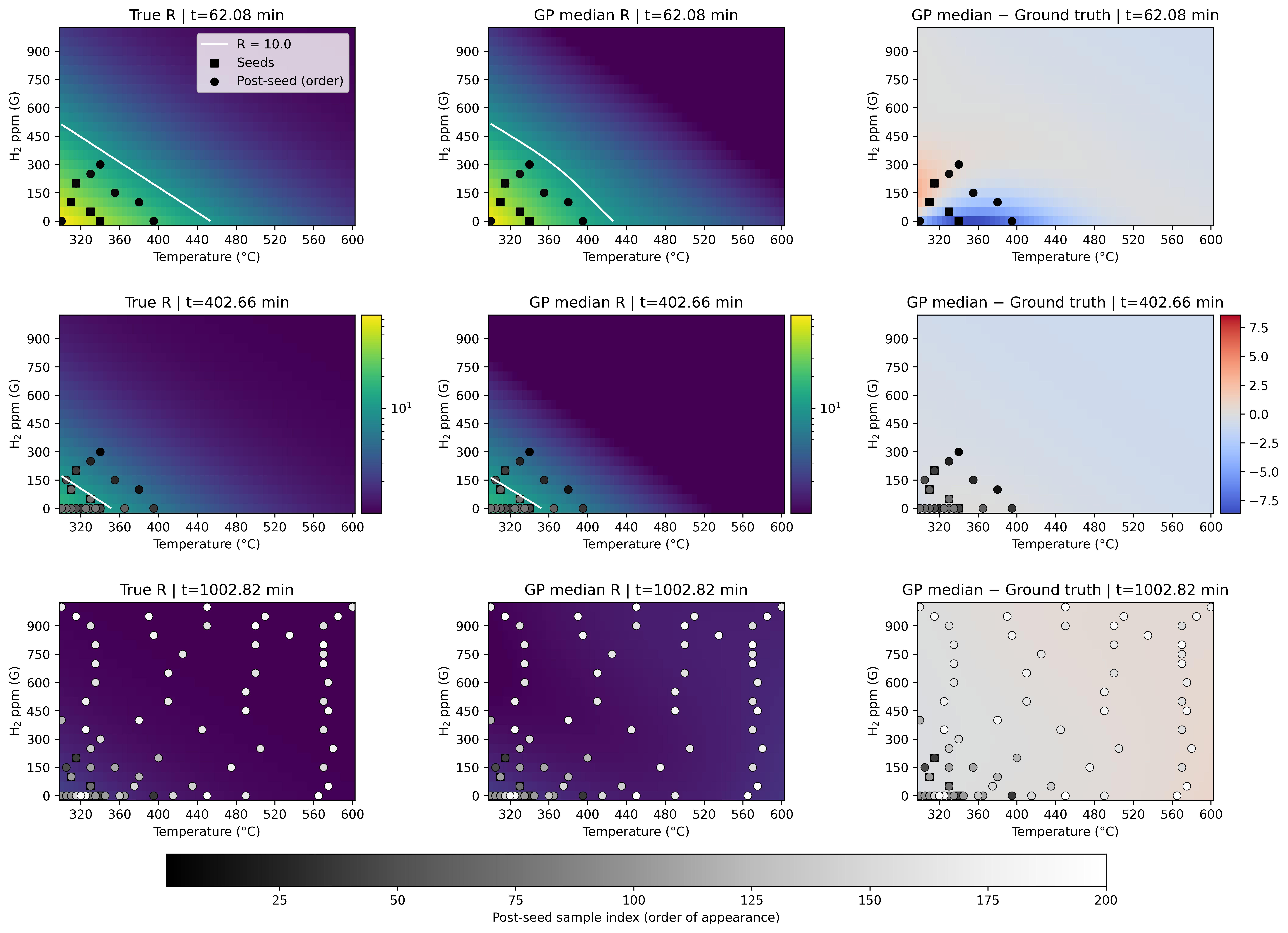}
    \caption*{Fig. S6. Evolution of the Gaussian process (GP) surrogate learned by SAL in the presence of measurement noise (10\% relative noise with a 0.5 absolute floor). Each row shows the rectification landscape at a different time during the simulated campaign. The left column shows the true rectification surface $R(T,G)$, the middle column shows the GP posterior median predicted by SAL, and the right column shows the difference between the GP prediction and the ground truth. Black squares denote the initial seed measurements, while circles correspond to samples collected during the SAL campaign (colored by acquisition order). Despite the presence of measurement noise, the GP surrogate progressively reconstructs the underlying rectification landscape as the campaign expands sampling across the domain. The residual maps show small, spatially smooth deviations across the domain, indicating that the GP estimates the underlying latent rectification surface rather than interpolating individual noisy observations.
    }
\label{fig:sal_snapshots_noise}
\end{figure*}

\begin{table*}[t]
\centering
\caption*{\textbf{Table S1.} Safe Active Learning (SAL) configuration used for the characterization of a  Pt/Cr\textsubscript{2}O\textsubscript{3}:Mg/$\beta$-Ga\textsubscript{2}O\textsubscript{3} device in this work.}
\label{tab:sal_params}
\footnotesize
\setlength{\tabcolsep}{4pt}
\renewcommand{\arraystretch}{0.9}
\begin{tabular}{lll}
\hline
\textbf{Category} & \textbf{Parameter} & \textbf{Value} \\
\hline

\multicolumn{3}{l}{\textit{Global Active Learning}} \\
\hline
 & $T_{\text{bounds}}$ & (350, 550) $^\circ$C \\
 & $G_{\text{bounds}}$ & (0, 800) ppm \\
 & Seed temperatures ($T_s$) & [350, 360, 370, 350] $^\circ$C \\
 & Seed gas concentrations ($G_s$) & [100, 0, 200, 150] ppm \\
 & Safety threshold & 150 \\
 & $\beta$ (exploration weight) & 3.5 \\
 & Safe fraction & 0.95 \\
 & Time horizon ($t_{\text{horizon}}$) & 40 h \\
 & Time step ($t_{\text{step}}$) & 2 h \\
 & Initial trust region $\Delta T$ & 25 $^\circ$C \\
 & Initial trust region $\Delta G$ & 100 ppm \\
 & Trust region warmup iterations & 5 \\
 & Trust region $\Delta T$ & 50 $^\circ$C \\
 & Trust region $\Delta G$ & 200 ppm \\
 & Max sequential failures & 5 \\

\hline
\multicolumn{3}{l}{\textit{Phase 1}} \\
\hline
 & Max iterations (Phase 1) & 100 \\

\hline
\multicolumn{3}{l}{\textit{Phase 2}} \\
\hline
 & Iterations (Phase 2) & 100 \\
 & $\tau_{\text{final}}$ & 1.0 \\
 & $\tau_{\text{half-life}}$ & 2.8 \\

\hline
\multicolumn{3}{l}{\textit{IV Sweep Parameters}} \\
\hline
 & Number of IV sweeps & 4 \\
 & Rectification voltage & 1.2 V \\
 & Measurement compliance & 0.1 A \\
 & Sweep delay & 10 s \\

\hline
\multicolumn{3}{l}{\textit{Device Equilibrium}} \\
\hline
 & Await device equilibrium & True \\
 & Device slope threshold & 1.0 \%/min \\

\hline
\multicolumn{3}{l}{\textit{Temperature Stabilization}} \\
\hline
 & Temperature tolerance & 2.0 $^\circ$C \\
 & Stability threshold & 2.0 $^\circ$C \\
 & Consecutive stable readings & 5 \\

\hline
\multicolumn{3}{l}{\textit{Gas Flow Control}} \\
\hline
 & Await flow stabilization & True \\
 & Flow tolerance & 25 ppm \\
 & Total flow & 1000 sccm \\
 & H$_2$ concentration & 5.0 \% \\
 & H$_2$ max flow & 100 sccm \\
 & Gas tolerance & 5.0 ppm \\

\hline
\end{tabular}
\end{table*}

\begin{figure*}
    \centering
    \includegraphics[width=\textwidth]{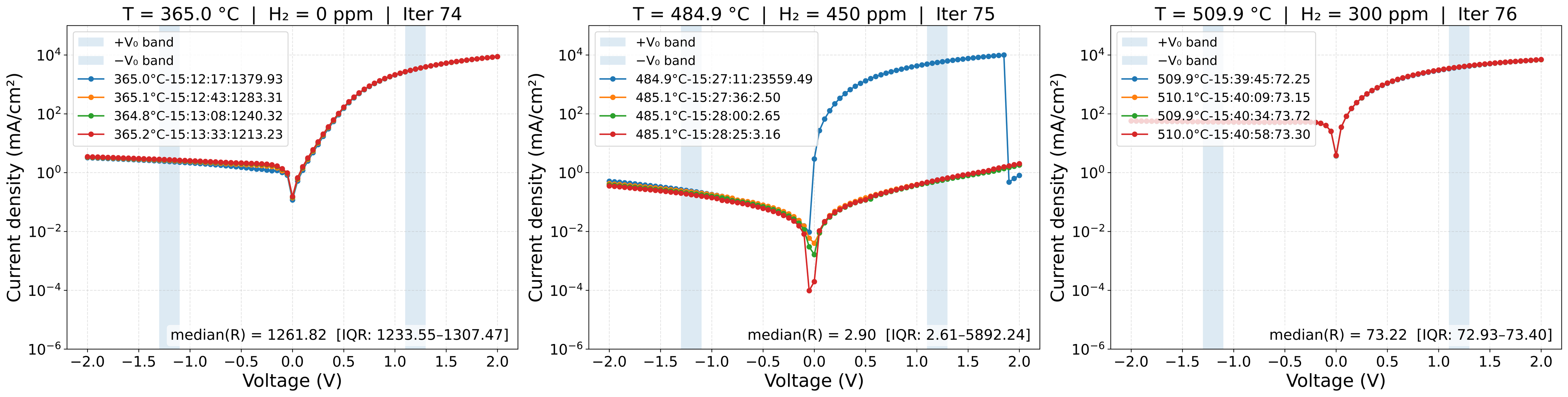}
    \caption*{Fig. S7. Bad measurement found during SAL, which resulted in an unsafe measurement at iteration 75 of the SAL campaign discussed in this work. Three out of four sweeps resulted in very low rectification ratios compared with the first sweep, which also showed measurements issues at high forward bias. Reliable measurements at iterations 74 and 76 suggest that the measurements at iteration 75 were not reliable and therefore were excluded from long-horizon forecasting.}
    \label{fig:bad_data}
\end{figure*}

\end{document}